\documentclass{sig-alternate}

\newif\ifsubmission
\submissionfalse % submission version

\usepackage{graphicx}
\usepackage{url}
\usepackage{amsmath}
\usepackage{times}
\usepackage{xspace}
\usepackage{multirow}
\usepackage{color}
\usepackage{caption}
\usepackage{subcaption}

\ifsubmission
\else
\fi

\newcommand{\comment}[1]
{\par {\bfseries #1 \par}} %comment showed

\newcommand{\liberouter}{\textit{Liberouter}\xspace}

\newcommand{\scampi}{\texttt{SCAMPI}\xspace}
\newcommand{\ibr}{\texttt{IBR-DTN}\xspace}

\newcommand{\peershare}{\textit{PeerShare}\xspace}

\newcommand{\ctracker}{\textit{Content Processor}\xspace}
\newcommand{\cgenerator}{\textit{Content Generator}\xspace}
\newcommand{\icomm}{\textit{ICN Communicator}\xspace}
\newcommand{\sandbox}{\textit{Sandbox}\xspace}
\newcommand{\wserver}{\textit{Web Server}\xspace}

\newcommand{\descr}[1]{\medskip\noindent \textbf{#1}}

\newcommand{\chroot}{\texttt{chroot}\xspace}
\newcommand{\seccomp}{\texttt{seccomp-bpf}\xspace}

\title{Bringing Modern Web Applications to Disconnected Networks}
\newcommand\Mark[1]{\textsuperscript#1}

\numberofauthors{1}
\author{
\alignauthor Marcin Nagy\Mark{1}, Teemu K\"arkk\"ainen\Mark{2}, Arseny Kurnikov\Mark{1}, J\"{o}rg Ott\Mark{2}\\
                \affaddr{\Mark{1}Aalto University, Finland, \texttt{firstname.lastname@aalto.fi}} \\
                \affaddr{\Mark{2}TU M\"unchen, Germany, \texttt{kaerkkae|ott@cs.tum.edu}} \\
\and
}

\ifsubmission
\else
\makeatletter
\let\@copyrightspace\relax
\makeatother
\fi

\ifsubmission
\else
\fi

\begin{document}

\ifsubmission
\else
\pagenumbering{arabic}
\thispagestyle{plain}
\fi

\maketitle

\begin{abstract}

Disconnected networks arise in environments where stable, well-connected
infrastructure connectivity is not ubiquitous.
In practice, these arise particularly in mobile systems, where the connectivity
can be achieved through short-range, pair-wise contacts between mobile
nodes.
Information-centric Networking provides a high-level framework
through which many different systems for disconnected networking can
be understood.
In this paper, we present a system built on top of the ICN abstraction that
allows interactive web applications to be developed and deployed in
disconnected networks.
We describe our system and protocol design, validate its operation
using simulations, and report on our implementation including adaptations
of six existing apps into web applications capable of being deployed in
disconnected networks.

\end{abstract}

\section{Introduction}
\label{sec.introduction}
%% Third attempt at selling this stuff...

%% 1. Establishing context

Information-centric Networking (ICN) is a paradigm that shifts the focus from
the communicating end-points to the content of the communication
instead~\cite{icn-survey}.
The argument for designing ICNs arises from the observation that in most
of the real world communication it is the content, or information, that is important,
not who is communicating with whom~\cite{icn-trossen}.
Therefore, ICNs name and operate on named data objects or information, rather
than on the network end-points.
This logically decouples the senders and receivers, which allows for natural
in-network caching (content can be stored anywhere) and multi-party communication
(content can be replicated to many receivers).
Using the ICN abstraction that more closely matches the reality of how the
communications are structured allows for more efficient networks to be
designed and deployed.
This in turn may enable novel services to be designed that would have been
too inefficient to be realized on end-point-centric architectures.

% - ICN technologies cannot abstract away the physical reality
% - when running on a well-connected physical infrastructure problems are easy
%		- depending on the solution, applications are more or less efficient, but not
%			do not fundamentally break (information gets from source to destination
%			in the expected time scale)
%	- when the physical infrastructure is not well-connected, we might use a different
%		dissemination strategy, e.g., direct contacts
%		- cannot just trivially adapt existing applications because the physical reality
%			will leak through the ICN abstraction as long delays between expressing the
%			desire to get some information and the information being delivered
%		- ICN platforms targeting this context exist (NetInf, Scampi, DTN routers)

What the ICN approach cannot do, is to abstract away the fundamental reality of the
underlying physical infrastructure from the applications running on top of it.
In well-connected physical infrastructures, it is possible to layer IP over ICN to
transparently support all the existing Internet applications and their implementations,
while gaining benefits from the intrinsic caching and multi-casting features of
ICNs~\cite{ip-over-icn}.
However, this requires the network to support low latency interactions on a global
scale, since that is the underlying assumption in the design of the transport and
application protocols on top of IP.
When physical reality breaks this assumption, e.g., when the network is
composed of short pair-wise contacts between nearby mobile nodes, {\em dissemination
strategy} of ICN must change.
Typically a space path oriented dissemination is replaced with a space-time path oriented one, or
in other words, packet switching is replaced with store-carry-forward dissemination. 
We refer to these as {\em disconnected} ICNs, which arise particularly frequently
in mobile systems.

The opportunistic and delay-tolerant networking communities have produced
many solutions to these scenarios, including
\textit{Haggle}~\cite{haggle-arch,haggle-impl,haggle-search-arch},
\textit{SCAMPI}~\cite{scampi-chants},
\textit{IBR-DTN}~\cite{ibr-dtn}, and
\textit{PodNet}~\cite{podnet,mobile-opp-ccn}.
These designs have tended to naturally converge towards the ICN principles,
but there are also designs explicitly starting from the ICN principles, such as
\textit{NetInf}~\cite{netinf}.
The underlying physical network may be composed purely of opportunistic
contacts between mobile nodes, or may include lightweight infrastructure,
or {\em throwboxes}~\cite{throwboxes}, such as the {\em Liberouter}
system~\cite{liberouter}.
Unlike well-connected ICNs, disconnected ICNs require explicit adaptations and
redesigns of the applications.
This is due to the complete break down of assumptions about reachability and
latency that are fundamental to classic application designs.

%% 2. How does the work advance state of the art

In this paper, we focus on bringing the most widely used classic
Internet application, the Web, into disconnected ICNs.
We can identify two distinct types of sites in today's Web: 1) content based
sites, and 2) web applications.
The former are sites composed of content that is static at any given point of
time.
Examples of these are company or institution websites, which rarely
change, and also websites with frequent content changes like news sites.
These static sites are easy to conceptually map onto ICN architectures, by treating
the entire site with all of its resources (HTML, CSS, JS, images) as content
items named by their URLs.
Bringing these types of web sites into disconnected ICNs has been
demonstrated in the past~\cite{dtn-wlan}.

However, many Web sites today are not composed of static content, but are
more accurately characterized as {\em Web Applications}.
These are sites designed to let users perform various functions, tasks and
activities, rather than simply presenting static content to them.
Many of the most popular Web sites of today fall into this category, e.g., social
networking (Facebook, Google+), photo sharing (Instagram, Snapchat), and
messaging (Twitter).
They are effectively equivalent to native applications, except they happen to
execute inside a Web browser context.

These types of web sites do not map semantically cleanly onto ICN architectures
the same way as the static content based sites do.
This is because a key feature of these web applications is reliance on back-end
services reachable via the Internet, and typically exposed via RESTful APIs.
The API interactions require stable and low-latency connectivity between the
browser and the back-end, which is only possible in well-connected networks.
%

%% 3. Detailed explanation of benefits

%% 4. Problems/challenges that must be tackled

%% 5. Solution

Our contribution in this paper is to present a system design and implementation
that enables web applications to be developed for, and deployed in, disconnected
ICNs.
The solution is based on leveraging three key concepts: 1) a {\em throwbox}
based physical network, and 2) the {\em caching}, and 3) the {\em multicast
dissemination} naturally performed by the disconnected ICNs.
While not every disconnected ICN design has these characteristics,
many practically deployable systems do.

In particular, we extend our previous throwbox design, the {\em Liberouters},
to include a framework that can serve interactive web applications to nearby
clients through standard, unmodified web browsers.
We show how to bundle together the {\em data} and the {\em logic} that comprises
the classic web application.
These bundles are then spread in the network through the caching and multicasting
mechanisms inherent in the operation of disconnected ICNs.
Any instance of our framework that receives a copy of a bundle can then use
contained data and logic to instantiate the web application accessible to locally
connected clients.
%
%The clients are not modified and can run unmodified end-user equipment with
%standard web browsers.

%% 6. Result overview
% - Show generality of our approach by using two different disconnected ICN platforms

Our results show that it is possible to adapt the traditional centralized web application
model to disconnected ICNs by combining pieces of content and the related logic in a
single message.
We further demonstrate the practicality and generality of our system design through
a real implementation on two different disconnected networking middlewares.
Finally, we show through simulations that the solution does not incur large enough
overheads that it would adversely impact the content dissemination done by ICN.

%% 7. Paper structure

Next, in Section~\ref{sec.model}, we describe our system model and
architecture at a high level, and provide a justification for our
message based approach from a distributed systems perspective.
We then lay out the detailed design of our framework in
Section~\ref{sec.app-framework}, and its proof-of-concept implementation,
including integration with two different disconnected ICN platforms, in
Sections~\ref{sec.impl}~and~\ref{sec:router-integration}.
We demonstrate the use of the platform by adapting a number of
existing applications, including the Google People Finder web application,
to use our framework in Section~\ref{sec.applications}.
Finally, we show a simulation based validation of our system design in
Section~\ref{sec.validation}, review the related work in Section~\ref{sec.related},
and conclude in Section~\ref{sec.conclusion}.

%\vfill
%\columnbreak
%% Teemu: We need a better name for this section...
\section{System Model}
\label{sec.model}

\begin{figure}
  \begin{center}
\ifsubmission
    \includegraphics[width=0.85\linewidth]{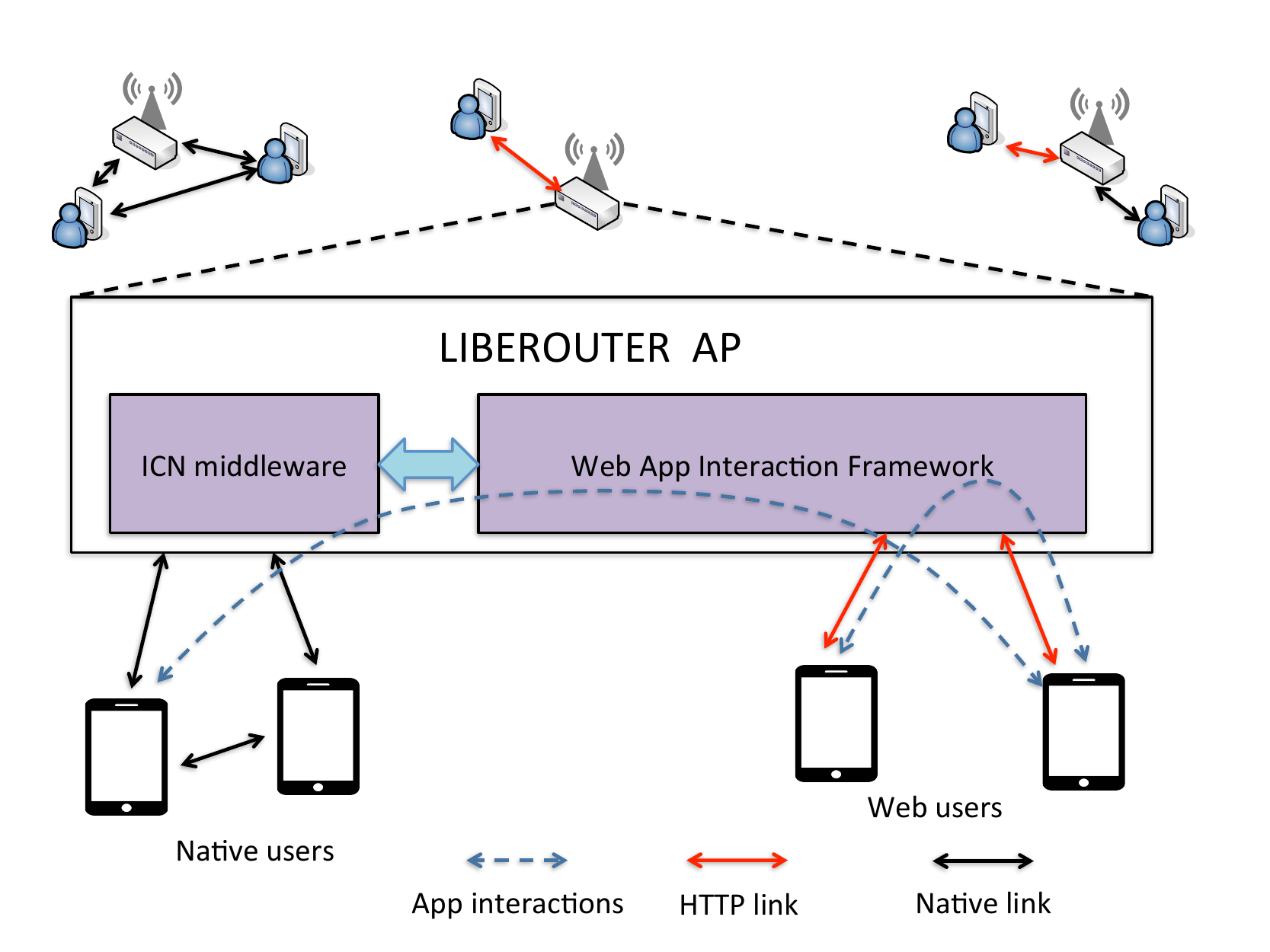}
\else
     \includegraphics[width=0.85\linewidth]{arch.pdf}
\fi
    \caption{Overview of the \liberouter based system model.}
    \label{fig.ecosystem}
  \end{center}
  \vspace{-3mm}
\end{figure}

%% From Section 3:
As described previously, our approach extends a \liberouter based
opportunistic networking system.
Figure~\ref{fig.ecosystem} shows the high level model of the system.
It is composed of opportunistic infrastructure nodes that act as access
points ({\em Liberouter AP} in the figure) to which nearby clients can
connect.
Our model considers two types of users: 1) {\em native users} whose
devices have native ICN support and applications
(bottom left), and 2) {\em web users} who are assumed
to only run an unmodified web browser (bottom right).
This leads to {\em native ICN} and {\em HTTP} links respectively.
The application interactions that we aim to enable through
our framework are between the web users and native ICN users.
We extend the functionality of the lightweight \liberouter devices by
adding a {\em Web App Interaction Framework} between a local
ICN networking layer and a web portal.
The interactions are bi-directional, allowing
the web users to both consume and produce content for the existing
ICN applications.

%% From previous intro
We will next explain the underlying messaging model from a distributed
systems perspective in Section~\ref{sec.messaging-model},
and show how our system can be understood from the applications'
viewpoint in Section~\ref{sec.app-model}.

\subsection{Messaging Model}
\label{sec.messaging-model}

\begin{figure}
  \begin{center}
\ifsubmission
    \includegraphics[width=0.6\linewidth]{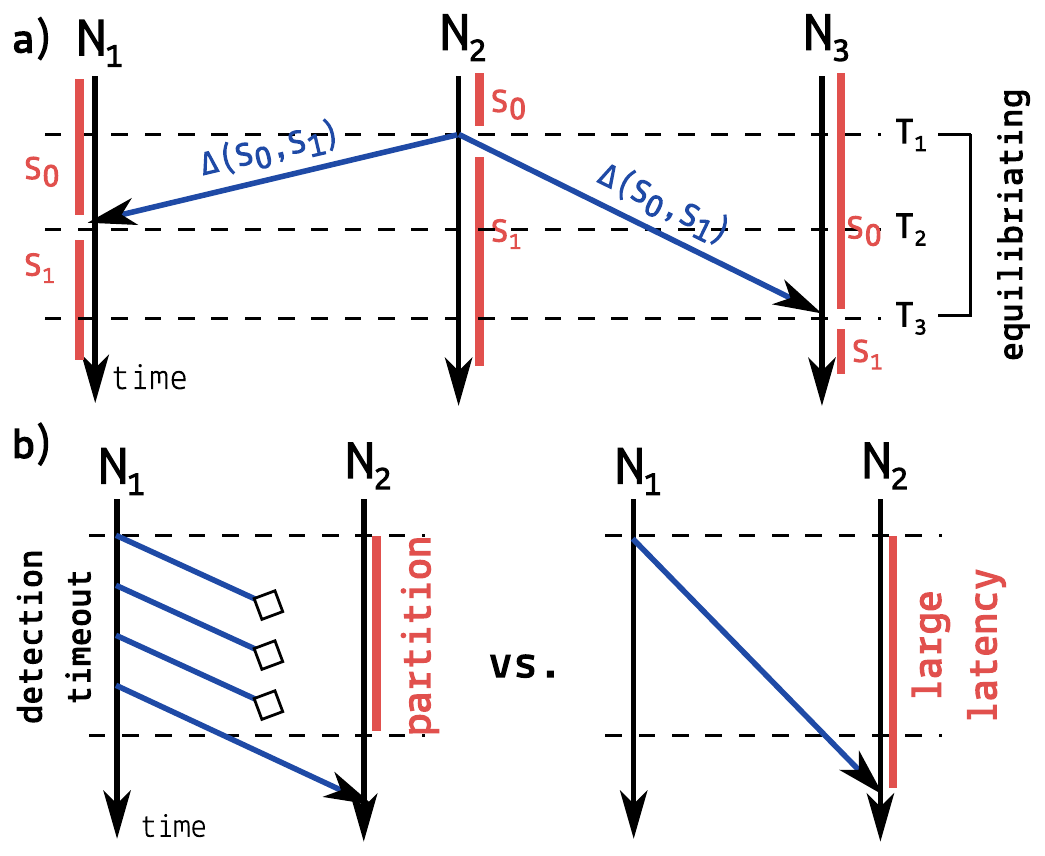}
\else
    \includegraphics[width=0.6\linewidth]{messaging-model-fig}
\fi
    \caption{a) Equilibration in a distributed system,
    b) Latency-partition duality.}
    \label{fig.messaging-model}
  \end{center}
\end{figure}

%% Distributed systems
The messaging model that we use can be derived from the fundamental
properties of distributed systems and opportunistic store-carry-forward
networks.
In general, distributed systems are designed to establish a globally
consistent state in multiple independent entities through message
exchanges.
This is shown in Figure~\ref{fig.messaging-model}a where a state
change $S_0 \rightarrow S_1$ in $N_2$ is propagated to the other
entities in the system through messaging.
This process takes place from time $T_1$ to $T_3$, which is the time
it takes for the message to propagate to all other entities.
During this time the system is in a globally inconsistent state, and
the process of reaching a consistent state is called {\em equilibration}.
The fundamental problem of distributed systems design is to ensure that
the equilibration process results in globally consistent states in the
face of concurrent state transitions in the entities.

%% CAP conjecture
The CAP conjecture~\cite{cap-theorem} is a useful tool for thinking
about the fundamental properties of distributed systems.
The major insight to be gained from CAP is that when facing
partitions (P), a distributed system design can trade availability (A)
against consistency (C).
Strict consistency can be enforced in some systems by a global locking
mechanisms, but it leads to no availability when the system is partitioned
and the lock cannot be acquired.
Consensus mechanisms employing quorums alleviate this by requiring only
a subset of the entities to agree on state changes.
This effectively leads to the larger part of a partitioned system to
remain available, at the expense of the other part having no availability
and an inconsistent view of the global state.
Mechanisms such as using soft state and caches closer to the clients
ensure availability in the case of partitions, but can lead to
a globally inconsistent state and problems when trying to reconcile
inconsistent states after a partition ends.
This essentially buys more availability at the cost of going from
strict consistency to eventual consistency.
In all cases, a key assumption in the design of both centralized and
peer-to-peer distributed systems is that partitions are transient
phenomena, and the system can eventually reach a consistent and
available state.
%

%\begin{figure}
%  \begin{center}
%    \includegraphics[width=6cm]{images/latency-partition}
%    \caption{Latency-partition duality.}
%    \label{fig.latency-partition}
%  \end{center}
%\end{figure}

%% Latency-partition duality
An important observation regarding the CAP formulation is that partitioning
is tightly coupled with latency, forming a type of latency-partition duality.
This is because in the absence of explicit knowledge about the network,
a partition is not distinguishable from a long delay, as illustrated in
Figure~\ref{fig.messaging-model}b.
This forces system designs to use delay thresholds as indicators for
partitions, which in turn makes implicit assumptions about the
information propagation speed in the network.
In particular, it assumes that latencies are in the order of user-acceptable
application level transaction times; an assumption that holds in well-connected
infrastructure networks where failures are transient conditions.

%% Opportunistic networks
Disconnected store-carry-forward networks are composed of pair-wise
contacts between mobile nodes.
This means that the information propagation latency is limited primarily
by the inter-contact times between the mobile nodes, and not by the speed
of light (and queueing) as in well-connected networks.
Combining this with the latency-partition duality, such systems can be
seen as being in a constant state of partitioning, causing the breakdown
of the underlying assumptions of the mechanisms used to maintain consistency
properties in ``classical'' distributed systems.
Another way to state this is that the time for a distributed system built
on such a network to equilibrate can be orders of magnitude larger (even
unbounded) than the time scales required by meaningful application semantics.
I.e., changes are made much faster than the time it
takes to reach an equilibrium.

This leads to the need to abandon the idea of global consistency, and instead
build distributed systems that are inconsistent by design\footnote{There are
inconsistent-by-design distributed systems built for well-connected networks
too, such as the git distributed source code management system.}.
A system that is inconsistent by design requires each entity to build their
own {\em locally consistent} view of the world based on their own set of
observations.
In concrete terms, it is the client software's responsibility to create a
locally consistent view or state from the (random) set of messages that it
has received.
Another way to state this is that the system enforces only local invariants.

%% Implications for messaging
Messaging is used as a way to transition the state of the entities in
distributed systems.
This can be written as $m: \Delta(S_x, S_{x+1})$, where $S$ is the
state of the system.
In other words, a message contains the difference between the two
states, which the recipient can apply to its state $S_x$ to transition
to the state $S_{x+1}$.
This has two major implications: 1) the recipient must be in the given
state $S_x$ or the message is useless, and 2) as a result of applying
the message, the recipient will be in the precise state $S_{x+1}$.
This makes sense in distributed systems that are designed to (periodically)
reach globally consistent states.

The above messaging model does not make sense in distributed systems
that are inconsistent by design, including opportunistic networks.
First, the messages should be applicable in any state, otherwise in a
system where every entity is potentially in a different state, most messages
would be useless.
Second, since the system is not designed to ever reach a globally consistent
state, the messages do not need to result in all the recipients moving to the
same state, just some locally consistent state.

This leads to a different messaging model, where each message will cause
a different state transition from one locally consistent state to another
in each receiving entity.
This can be written as $N(m): S_N \rightarrow S_{N'}$, where node $N$
applies message $m$ to transition from the locally consistent state $S_N$
to another locally consistent state $S_{N'}$.
We can observe that it must be possible to apply the transition from
any state, including the {\em empty} state, into some locally consistent
state; $N(m): \emptyset \rightarrow S_N$.
This is typically expressed in opportunistic and delay-tolerant networking
as messages being {\em self-contained} and {\em semantically meaningful}.

It is this fundamental property of self-contained messages that we exploit in
building our framework.
This messaging model further implies that the network will have a large
number of these messages (i.e., content), which can be interpreted
independently of any specific application state.
However, the applications themselves are still required in order to
participate in the system.
The key idea of our framework is to ship minimal, generic application
logic along with the messages, which allows any node a degree of interaction
with the system (e.g., view, respond and send) without requiring the
specific native application.

%It should be noted, that while the messages are independent of client
%states, there may be inter-dependencies between messages.
%%
%For example, one message can be a {\em reply} to a previous one.
%%
%This does not pose much of a problem, since the full dependency
%chain of messages can be included in a transport unit by the
%sender.

%% Application model through MVC

\subsection{Application Model}
\label{sec.app-model}

\begin{figure}
  \begin{center}
\ifsubmission
    \includegraphics[width=8.5cm]{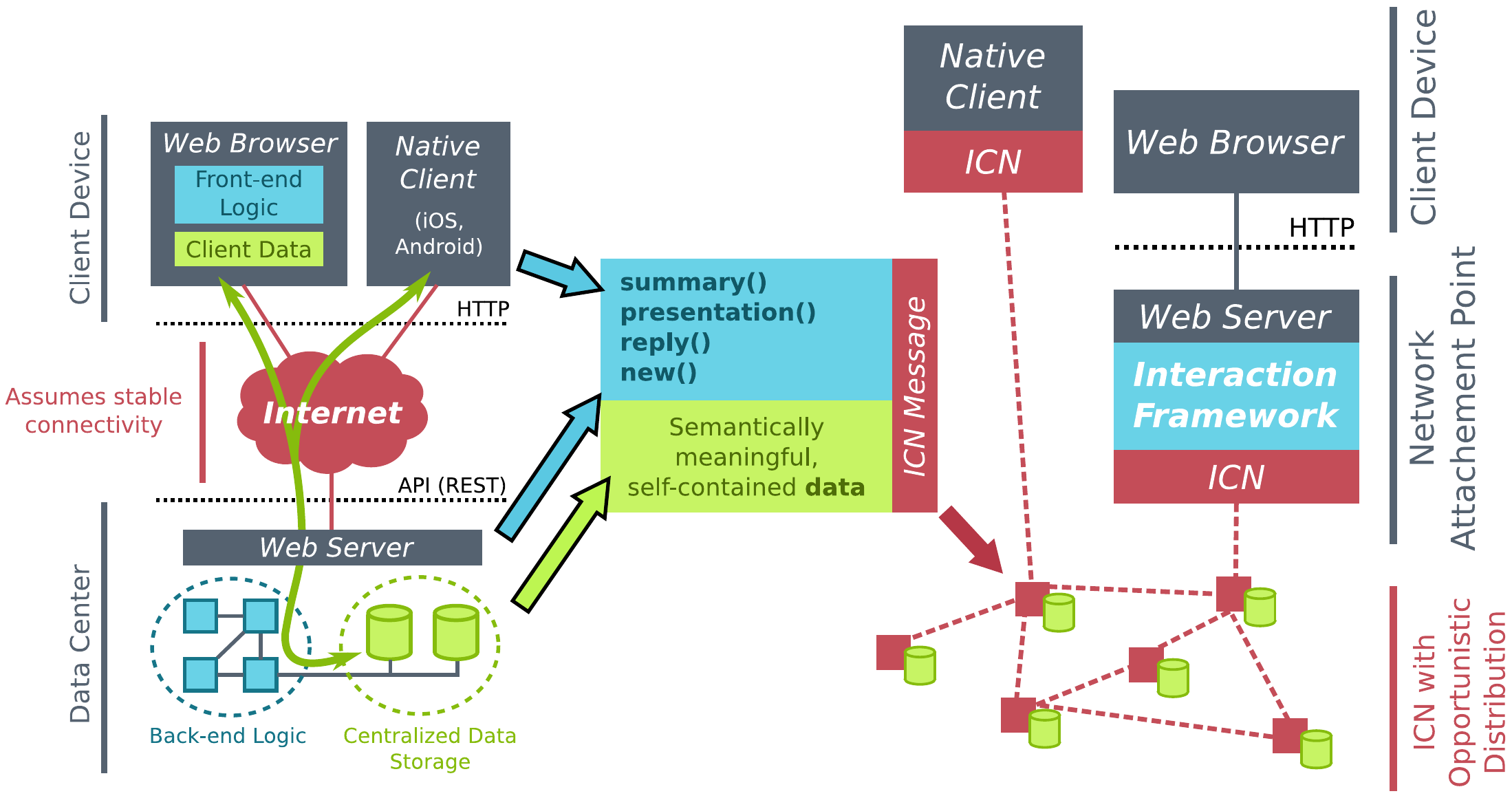}
\else
    \includegraphics[width=8.5cm]{web-system-model}
\fi
    \caption{Bundling content and logic to enable distribution in a disconnected ICN.
    \vspace{-5mm}}
    \label{fig.web-system-model}
  \end{center}
\end{figure}

Traditional web applications are built around a centralized client-server model, as
shown on the left of Figure~\ref{fig.web-system-model}.
In this model, all application data are stored in the central server, along with the
back-end logic that controls interactions with that data.
The server exposes an interface towards the clients, typically through a RESTful
API.
The clients run the web application front-end logic inside a browser as a collection
of JavaScript libraries and code, which calls the back-end services through the API.
The front-end logic fetches the relevant pieces of data from the back-end for presentation
to the user, and sends back the user's data for storage in the central database and for
sharing with other clients.

The fundamental underlying assumption in the client-server model is the existence
of a stable and fast network connecting the clients to the server.
This assumption is satisfied by the Internet in most of existing deployments, and can be
equally satisfied by a well-connected ICN.
However, ICNs can also be deployed in disconnected environments, where the
dissemination strategy is build around exploiting short-lived one-to-one contacts
between devices.
In such a dissemination mode, the client-server approach becomes impractical, as it is
generally not possible to support the low-latency end-to-end interactions required by
the back-end API.

To enable web-browser based applications in disconnected ICNs, as
shown in the right of Figure~\ref{fig.web-system-model}, we need to deal with two fundamental aspects of the web applications:
{\em data} and {\em logic}.

First, the {\em data} of the application cannot be stored in a centralized database,
because reaching such a database with a reasonable latency is generally not possible.
Solving this in ICN is straight forward since the originator of the data can simply
publish content into the network as self-contained, semantically meaningful messages,
which become accessible to other nodes without the need for a centralized database.
For example, a social networking application does not need to get an image posted by
the user sent to a data center on another continent, but can simply disseminate it directly
to nearby peers.
Consequently, we go from a centralized database into a distributed storage of pieces of application
data using the resources available directly in the disconnected ICN.

Second, just getting access to data does not solve the whole issue.
We must also be able to run {\em logic} to generate presentation of data to display to
the user, and to generate new data into the network in response to user actions.
We solve this by attaching the logic to each message along with the data.
This ensures that data and logic essentially share fate -- If there is no data available,
then there is no need for the application logic to present the data; if there is no application
available, then there is no value in having data since it cannot be presented.
This bundling of data and logic is shown in the middle of Figure~\ref{fig.web-system-model}.

Attaching the full front-end application and related back-end logic to every piece of application
data would result in large overheads.
We therefore create an abstraction of the application logic which is split into two parts:
1) a {\em generic application model}, which defines the states and transitions of the application, and
which is implemented by the \liberouter device running the framework.
In this paper we use a simple summary/detail model where the application has two main states
or views; a {\em list of summaries} of each message, where the user can select a particular
message to get a {\em detail} view of that message.
Different generic models could be defined and each message could indicate which model is
the best fit to present it.
2) specific {\em functions} that are executed at different points in the generic application flow to
generate the particular view of, or the action on, a message.
This is the logic that is carried with each message and which is defined by the application
developer as scripts.

\section{Web App Interaction Framework}
\label{sec.app-framework}

%\begin{figure}
%  \begin{center}
%    \includegraphics[trim=2cm 2cm 6cm 4.5cm,width=6cm]{web-if}
%    \caption{Interfacing message-based communication applications to
%    legacy devices using web technology.}
%%    non-liberouter devices using HTTP, HTML, and JavaScript.}
%    \label{fig.web-if}
%  \end{center}
%  \vspace{-0.7cm}
%\end{figure}
%
%In this section, we address the interaction of the \liberouter
%framework with \textit{legacy} devices that do not run the \liberouter
%code.
%%
%The basic idea is to implement a set of web-based extensions for the captive portals of \liberouter APs.
%
%%%Figure~\ref{fig.web-if} illustrates these extensions. We leverage the fact that the \scampi router stores encoded messages in the \liberouter file system (1a) to provide two data access paths: one, (8a) and (8b), allows for message forwarding (see section~\ref{sec.web-forwarding}), while the other, (2)--(4) and (5)--(7), allows for stored content access and upload (see section~\ref{sec.web-content}).

We now turn to a detailed description of the framework.
We begin with the description of the framework and its functionality at a conceptual level (Section~\ref{sec:conceptual-model}).
Extending message with presentation and interaction logic transformations by shipping them as meta data is explained in Section~\ref{sec:meta-data}.
%
%Among various meta data we have defined four special types that carry
%application specific embedded scripts (see section~\ref{sec:scripts}),
%and are used for generation of HTML code (used for presentation of message
%content), and allows for potential new content generation inside the web
%browser.
%
Section~\ref{sec:framework-design} provides a detailed description of the framework including functionality and interactions between framework components. Section~\ref{sec.app-dist} describes the process of bootstrapping web applications. Finally in Section~\ref{sec:security}, we discuss security implication for the framework design.

\subsection{Conceptual Model}
\label{sec:conceptual-model}

Recall from Section~\ref{sec.model} that self-contained and semantically meaningful messages are the foundation of our framework design. 
Our goals are to: (1) enable {\em generation} and {\em presentation}
of content by web browsers through the framework and (2) provide
interoperability between native applications and their web
equivalents.
We could easily achieve content presentation by embedding HTML describing content into a message. But this would not
allow for generation of new content and would tightly couple the application
state with the message interpretation. %It might also imply replicating
%(part of) the content, natively and as HTML.
%
Therefore, we choose a more general approach. In the rest of paper, we use {\em message} to refer to the transport unit in an underlying ICN and {\em content} as general data associated with an application.
%Therefore, we model the interpretation and generation of messages
%as transformations on opportunistic messages.

%Conceptually we model opportunistic {\em applications} and {\em messages}
%belonging to these applications.
%%
%Each message is assumed to be self-contained and semantically meaningful.
%%
%The applications have native clients that are the primary means for
%generating and consuming the messages.
%%
%Our goal is to enable web {\em generation} and {\em consumption} of the messages
%by web browsers through the \framework.
%%
%The consumption goal can be trivially realized by embedding HTML code that describes content of the message.
%%
%However, this approach does not allow for seamless message exchange between opp and legacy users.
%%
%Additionally, it also does not separate application state and its presentation making it difficult for app developers.
%%
%Thus, we model the consumption and generation of messages as transformations of a set of opportunistic messages.

The interpretation of content can be seen as a transformation of
a set of messages into a view rendered by a web browser, i.e.,
generation of an HTML page from a set of messages.
There are two aspects to this transformation: 1) an individual message
must be transformed into an HTML view, and 2) these individual
views must be composed into a coherent application view.
For example, a photo sharing application would be composed of a transformation
of individual messages into, e.g., thumbnail views, and a composition
of those thumbnails into an application view, e.g., list sorted by the date.

For the first aspect, we define two transformations: {\em message summary} and
{\em message presentation}.
The {\em message summary} transformation generates a concise view of the message,
suitable for inclusion in a listing of a large number of messages.
The {\em message presentation} transformation generates a detailed view of the message,
intended to be displayed on its own to a user who is interested in the
message content.
Both transformations are pure functions that take as input the message
and generate the presentation view as output (i.e., HTML code).
The transformations are message type and application dependent, and are
expected to be included in the message itself by the original generating
application.

As messages are self-contained,
% and semantically meaningful,
the message transformations alone already provide a useful
view into the messages.
However, in many applications there may be further requirements on the
display of sets of messages.
For example, a message board application should list its messages
in time order and possibly threaded by a topic.
This corresponds to an application level logic which would normally be
implemented by the native clients.
%
%While our immediate goal is not to enable fully fledged web applications to be
%built on top of the framework, it must be possible to apply some degree of
%application layer composition logic in order for the generated views to
%be meaningful.
%
To this end we define {\em application presenter} transformation.

\begin{figure}
  \begin{center}
\ifsubmission
    \includegraphics[width=\linewidth]{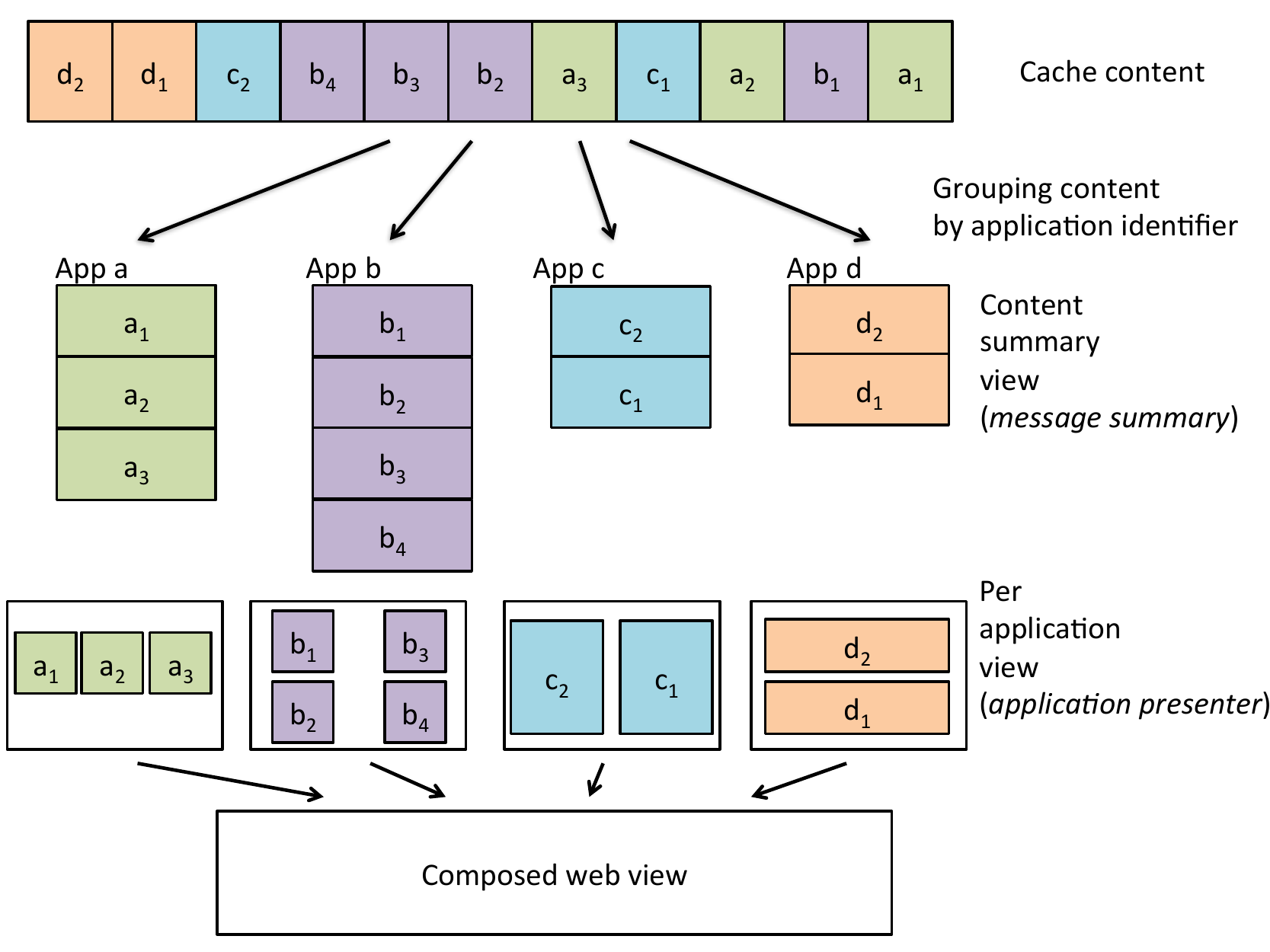}
\else
    \includegraphics[width=\linewidth]{message-processing.pdf}
\fi
    \caption{Processing of messages for content extraction and presentation in web browsers.}
    \label{fig.msg-processing}
  \end{center}
\end{figure}

The {\em application presenter} transformation is similar to the message
transformation in that it is a function that maps a set of
inputs into an HTML view to be displayed via a web browser.
As input, they take a set of messages (e.g., all messages belonging to
a particular application), and optionally some set of state generated
by a previously run presenter.
For example, a forum application could have one {\em application presenter} transformation that
processes all forum messages and generates a list of topics.
When the user selects a topic, another {\em application presenter} transformation would generate
a list of all messages within the topic.
In general, each application can have an arbitrary number of linked {\em application presenter}
transformations that can call each other. Figure~\ref{fig.msg-processing} illustrates the whole process of
building complex application level web views based on transformations included in application messages.

The generation of messages can take two forms: generating {\em new}
messages, or {\em replying} to an existing message.  % (which may itself be either new message or a response to an earlier message).
Both take as input a predefined set of values and compose a message, which
is then injected into the network by the framework.
The difference is that the {\em reply} transformation also gets as a parameter the
message to which the user is responding.

\vfill
\subsection{Shipping App Logic along with Messages}
\label{sec:meta-data}
To enable bundling of presentation and interaction logic with messages, we propose to attach transformations to messages as meta data in the key-value format. We define five basic keys, and each key corresponds to the specific transformation. The {\em appSummary} key contains the application presenter transformation. The message level transformations are contained in {\em summary} and {\em presentation} keys. Finally, for generating new content, {\em new} and {\em reply} keys carry appropriate transformations.

%Opportunistic applications use the Bundle Protocol~\cite{bundle-spec} as the basic communication protocol. We aim to embed transformations into messages without disrupting opportunistic network functionality for devices that do not understand embedded transformations. To this end, we have developed a simple {\em Web Message Interaction Format} (WMIF) that attaches transformations to messages as metadata.

%As a message carries arbitrary type of content, each transformation is assigned to a particular message. Thus, to enable it, we have developed a simple {\em Web Message Interaction Protocol} (WMIP) that carries transformations together with some additional data inside a message as a metadata. As Bundle Protocol supports extensions via metadata, WMIP can be easily added to an opportunistic network without disrupting operation of devices that do not support it.

%WMIF has key-value format, defines five basic keys, and each key corresponds to the specific transformation. The {\em appPresenter} key contains application presenter transformation. Message level transformations are contained in {\em summary} and {\em presentation} keys. Finally, for generating new content, {\em new} and {\em response} keys carry appropriate transformations.

There are also additional keys that help the framework to present content in the web browser. The {\em contentType} key indicates type of data contained in the message (e.g., if an item carries a photo, contentType should have ``image'' value). 
For simple items comprising only data of one type, the framework may take advantage of the \emph{contentType} key, and provide a simplified interpretation of the message content (e.g., by taking a thumbnail for the ``image'' \emph{contentType}). \ifsubmission More details on this feature are included in our tech report \cite{dtn-legacy-app-framework-full}. \else \fi
Other keys defined are: {\em description}, {\em service}, {\em icon}. The \emph{description} provides a short text summary of the message content (e.g., short comment on the location of photo shot). The \emph{service} key maps the message to the application name and \emph{icon} contains an application icon.
\ifsubmission 
\else 
Table~\ref{tab.meta-summary} provides an overview of defined meta data items. 
\fi

\ifsubmission
\else
\begin{table}[tb!]
  \centering
  \caption{Web Message Interaction Format summary}
  \resizebox{\columnwidth}{!}{%
    \begin{tabular}{|l|l|}
	\hline
      \textbf{Meta data key} & \textbf{Description} \\
      \hline
	contentType &  Indicates content type carried inside the message \\ 
	& (see table~\ref{tab.inferred_html} for list of available values) \\ \hline
	description & Text describing message content \\ \hline
	service & Application name \\ \hline
	system & Integrated software name \\ \hline
	icon & File containing application icon \\ \hline
    \end{tabular}	
  }
  \label{tab.meta-summary}
\end{table}
\fi

\begin{figure}
  \begin{center}
\ifsubmission
    \includegraphics[width=\linewidth]{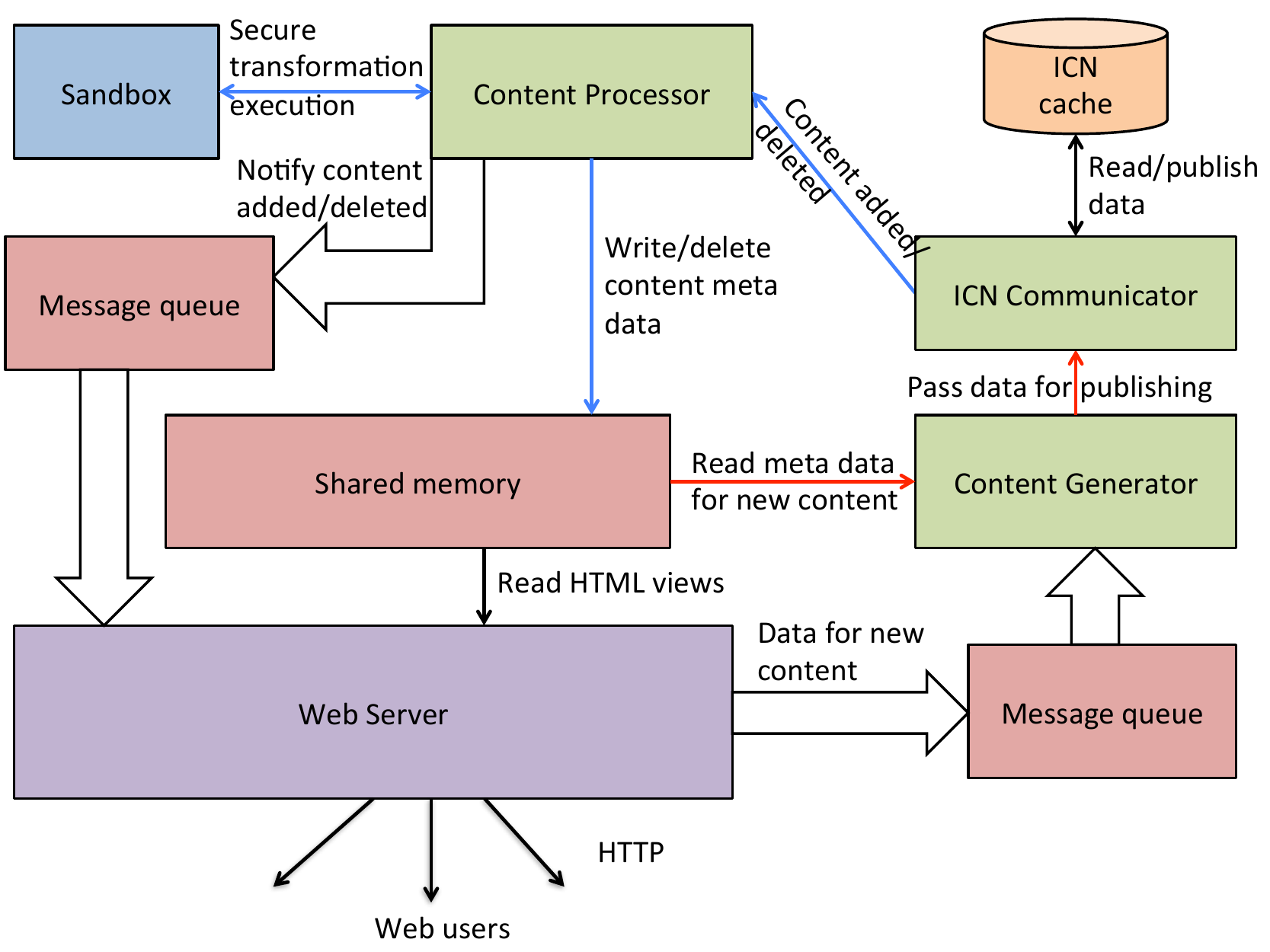}
\else
    \includegraphics[width=0.85\linewidth]{system-model-3.pdf}
\fi
    %\vspace{-3cm}
    \caption{System components and their interaction.}
    \label{fig.system}
  \end{center}
  %\vspace{-3mm}
\end{figure}

%\vfill
\subsection{Framework Design}
\label{sec:framework-design}
%The framework comprises four main components, namely \pgenerator, \presenter, \webhandler and \cgenerator. The first two units are responsible for interpretation of messages, and their arrangement as a part of HTML view. The two latter components enable web-based generation of new messages.
The framework comprises four main components, namely \ctracker, \cgenerator, \wserver and \sandbox. It also uses a message queue and a shared memory for inter-process communication. Finally, to allow for independence of the framework from an underlying ICN system, the framework defines also \icomm which (1) tracks changes of messages in the ICN cache and (2) acts as an adapter of message format for the underlying system. Figure~\ref{fig.system} shows an overview of the framework architecture.

%
% are used for obtaining external data provided by the user to respond to already existing messages, or create a completely new message. 

%Furthermore, to provide more information about the message, all Bundle 
%Protocol header fields are also treated as metadata, and made available for 
%the framework.

%%
%Finally, since an application may require more data to provide a usable content 
%presentation, the metadata protocol allows arbitrary metadata values, and makes 
%them available for the message interaction scripts to present message content or 
%to generate a new message.

\ctracker is notified by \icomm about changes in the ICN cache. If a new message is reported by \icomm, \ctracker extracts meta data contained inside the message and writes them into the shared memory. After that it checks if extracted meta data contain transformations and if yes, it asks \sandbox to securely execute all contained transformations and writes their outputs into the shared memory. Details of \sandbox design are explained in the Section~\ref{sec:security}. Execution of all transformations at this stage allows to have message presentation data cached if the web user wants to see them in the future, greatly improving his user experience. However, if meta data do not contain transformations, \ctracker provides a simplified presentation of message content, which is based on the value of the \emph{contentType} key. \ifsubmission \else Table~\ref{tab.inferred_html} describes mappings between the \emph{contentType} value and the HTML tag for simplified content presentation.\fi Finally, \ctracker notifies \wserver via the message queue about the new message that has appeared in the ICN cache by giving \wserver pointer in the shared memory to the item's data for access. Similarly, if removal of an existing message is reported by \icomm, \ctracker removes its details from the shared memory and notifies \wserver about it.

%In addition, the \ctracker is also an essential part of an app store provided by the framework. If a received message has bundleType value of ``app", the \pgenerator extracts binary contained in the message and makes it accessible for download via the web browser (see Figure~\ref{fig.webportal-screenshot}). %Obviously such a message may also contain transformations

\wserver is responsible for: (1) presenting stored message content to the user, and (2) handling requests to create new content coming from the user. It realizes message content presentation by implementing the generic web application model. It is build of a set of generic template views, which are filled with appropriate transformation outputs depending on type of view presented to the user. For example, if user wants to see whole content belonging to a particular application, \wserver fills template with the output of the \emph{application presenter} transformation. For a request to generate new content, \wserver presents user with an HTML form (result of the \emph{new}/\emph{reply} transformation) that has to be filled in with new item details. The user submits the form to \wserver which validates it, and if correct it passes it to \cgenerator via the message queue.

\ifsubmission
\else
\begin{table}[tb!]
  \centering
  \caption{Inferred HTML tags for given \emph{contentType} values}
  \resizebox{0.8\columnwidth}{!}{%
    \begin{tabular}{|l|l|}
	\hline
      \textbf{\emph{contentType}} & \textbf{Inferred HTML presentation} \\
      \hline
	audio & \textless audio \textgreater \\ \hline
	video & \textless video \textgreater \\ \hline
	image & thumbnail \\ \hline
	text & text beginning \\ \hline
	app & link \\ \hline
	other... & link \\ \hline
    \end{tabular}	
  }
  \label{tab.inferred_html}
\end{table}
\fi

\cgenerator component listens for new content request from the message queue. Upon its reception, it creates a new message based on data from the form and the shared memory, after that it moves the newly created item into the ICN cache. 

%The \cgenerator component listens for incoming content generation requests that are sent by the 
%\webhandler. Upon reception of such a request, the \cgenerator 
%creates a new message based on data received from the \webhandler, and moves the 
%newly created message to the local router database (step (8) in
%Figure~\ref{fig.system}), so that it can be inserted into the network. 

%\subsection{Enhancing opportunistic applications for legacy users}
%\label{sec:extending-apps}
%To enhance an opportunistic application for legacy users access, an app developer must implement transformations he intends to use in the application. Among all possible transformations, only the application presenter transformation is the mandatory one. Implemented transformations must be embedded into application messages as meta data. 
%
%To facilitate the development process, we have also implemented a simple transformation testing environment. It allows the app developer to test correctness of his implementation by: (1) executing the transformation in it and (2) verifying that HTML view generated by the transformation is coherent with the application design.

\subsection{Bootstrapping Applications and Nodes}
\label{sec.app-dist}

So far, in our design, transformations are always shipped along with
the message. This ensures compatibility between content and the application logic.
However, it also implies that web users can generate content
for a web application only if \liberouter they connect to has already
stored a message of this application. In practice, this means that every web application
must be bootstrapped from its native equivalent. 
% even if the users wants to create
%a \emph{new} message---and this first message would need to be created
%by a native application.
%
To overcome this limitation, we take two further steps: (1)
distribute native applications within the framework to increase availability of native applications in ICN,
%transform legacy
%into opp nodes
and, more importantly, (2) allow creation of initial
content from the web applications.

Recall that a native application is essentially a structured binary content
(i.e., an executable plus auxiliary data).  As such, it can be wrapped
up inside a message and distributed across ICN to
become accessible via a disconnected ``app store''~\cite{liberouter}.  %This
%already makes ICN self-sufficient, except for the bootstrap
%problem that the native opportunistic router (e.g., SCAMPI or IBR-DTN)
%and the ``app store'' would need to be installed in the first place.

To address (1), the framework offers the disconnected ``app store'' functionality by making all native applications available
to web users for download via the web browser -- if a received message has \emph{contentType} value of ``app'',
\ctracker extracts an executable contained inside it and puts it inside the \wserver public directory.
 \ifsubmission. \else
(see Figure~\ref{fig.webportal-screenshot}). \fi
For (2), we also include transformations in messages used to
distribute native applications.  A user visiting \liberouter can
then not just choose to install a native version of an application, but also to
create new content using just a web version of the application.
%message for the respective app by just using the framework.

\subsection{Security considerations}
\label{sec:security}
%Our security considerations cover \ifsubmission two \else three \fi issues: (1) secure execution of transformations \ifsubmission and \else,\fi (2) message authenticity\ifsubmission\else and (3) providing access to encrypted content for legacy nodes\fi.

To guarantee robust and secure operation of the framework, our security considerations focus on four issues: (1) secure execution of a transformation, (2) message authenticity, (3) encrypted message access and (4) impact of communication mode on privacy. The first requirement is the most important one, as insecure execution of a malicious transformation may result in framework operation malfunction and unauthorized access to ICN cache. In a disconnected environment, access to an authentication server cannot be taken for granted, thus the framework must provide functionality for verification of message authenticity. \ifsubmission In our tech report~\cite{dtn-legacy-app-framework-full} we provide more details on privacy and accessing encrypted content. \else  \fi

\descr{Threat model and assumptions.} We are concerned with threats that an adversary: (1) disrupts framework/ICN functionality by executing a malicious code and (2) impersonates another user by sending a message that masquerades its originator for another user. Our threat model assumes also: (1) presence of basic Linux platform security mechanisms and (2) presence of a disconnected public key distribution (PKI) system~\cite{dtn-pki} which assigns a public key to an identifier of message originator in ICN. Examples of such PKI systems are SocialKeys~\cite{socialkeys} and \peershare~\cite{peershare}.

\descr{Secure execution of transformations.}
Execution of transformations of unknown origin poses a serious threat to the secure functioning of the framework. The malicious transformation may include system calls causing disruption of the framework operation and possibly even whole ICN (e.g., the transformation switches off all network interfaces). The other set of threats comes from unauthorized content access. They include pollution of content by generation of fake messages, deletion of other messages and access to information contained in other messages. This threat model motivates implementation of the file system level isolation and the system call filtering. File system level isolation constrains the transformation to access only data contained inside the message and parts of the shared memory that are related to it. As system calls are still available to the transformation (via system libraries), the system call filtering must prevent invocation of any system calls other than I/O operations on message content to which the transformation belongs (see section~\ref{sec.impl} for details). Since transformation access to file system is constrained and it cannot invoke system calls other than file system I/O, security requirements of transformation execution are fulfilled.  

\descr{Message authenticity.}
We address the threat of another user impersonation by providing mechanisms to verify message authenticity. It requires that message is signed by its originator. Verification of message authenticity can be realized on the framework side, or in the web browser. For the former, the framework itself verifies the signature of the message using the public key available in the ICN cache. The browser side verification assumes that the public key is accessible in the browser for the web application (e.g., it is present in HTML5 local storage). Section~\ref{sec.impl} provides implementation details.

\ifsubmission
\else
\descr{Encrypted message access.\footnote{Implementation of this part is still ongoing.}}
In the framework design, we make the assumption that content carried inside messages is unencrypted, thus easily accessible by the framework. However, we argue that web users experience can be further enriched by providing them also access to encrypted content. To do this, we assume availability of cryptographic keys in web users browsers which are either delivered via some key distribution system, or derived by web users by means of Password-based Cryptography~\cite{rfc2898}. Since encrypted content should be accessible only to users that have an appropriate key, it must be decrypted inside the browser (using either Web Cryptography API, or a specialized JavaScript module fetched from the framework). Finally, the transformations of decrypted messages must also be executed inside the browser (thus they must be written in JavaScript), so that they can be directly displayed to the web user without the necessity to interact with the framework. 
\fi

\descr{Communication mode and privacy considerations.}
Current framework design assumes that all messages stored in the ICN cache should be accessible for all users. 
This assumption holds well for all applications in which users do not target their
content at a (closed set of) recipient(s), but rather share it
openly. Thus, all users can read all messages for the group
using the framework. On the other hand, for applications targeting their content
at specific recipients, the framework must not give access to their messages 
to the unauthorized users, as it would violate user's privacy. The same problem holds 
for granting users access to encrypted messages. 
\ifsubmission
We leave this problem as future work and our tech report~\cite{dtn-legacy-app-framework-full}
provides more details.

%On the other hand, there are also opportunistic
%applications (e.g., Whisper) assuming a closed communication model. In
%such a case, the framework must give access to the message for the
%legacy user who is not the recipient of the message, as it would
%violate user privacy rules. 
\else
For these group of applications, the
current framework needs to be modified to support addressing scheme
for legacy users. This feature can be realised by generating a random
endpoint identifier on the first connection to the framework and
storing it persistently in the web browser as a cookie. As a result,
the framework may use cookie to identify the legacy user and grant
access only to his/her private messages.
\fi

%Executing scripts of unknown content to generate message presentation data, or 
%uploading arbitrary data to the system raises serious security concerns.
%%
%To address this issue, the framework implementation must execute the
%scripts in their own dedicated context with no access rights to anything 
%except for the message self-contained data.
%%
%In addition, to further prevent transformations from doing any harmful operations, the 
%framework must accept only such a subset of programming language functions that do not pose a threat to the system.
%%
%To do this, the framework could either provide its own language interpreter~\cite{OBC}, or run static analysis on the script 
%to detect potentially harmful code.

%chroot environment, seccomp

%% Teemu: AFAIK static analysis is heuristic based and doesn't provide
%% strict guarantees of correctness, or did you mean some theorem proving
%% system?

\section{Implementation}
\label{sec.impl}
We now turn to the implementation details. We first cover implementation of core framework components, followed by security components details, which are described separately due to their high complexity. After that we explain web application transformations and finally conclude this part with performance evaluation of the framework. 

% and its integration with \scampi and \ibr routers.
%Of \scampi open sourced applications, we provide web version of {\em GuerrillaPics}, {\em GuerrillaTags}, and {\em PeopleFinder}, while of \ibr applications, we extend {\em Whisper}, {\em Talkie}, and {\em Sharebox}.

%\subsection{Framework}
\descr{Core framework.} We implemented the framework as the combination of four applications, namely \ctracker, \sandbox, \wserver and \cgenerator. \ctracker component is developed in standard Java. It monitors updates to the ICN cache by: (1) parsing meta data of newly received messages, and storing them as a hash map inside Redis\footnote{Redis: http://redis.io/} (acting as the shared memory), and (2) clearing Redis of data removed from the cache. Textual meta data objects included inside messages are stored directly in Redis, while binary objects are written to the framework persistent storage, and only their access paths are stored in Redis. \ctracker uses Redis queue to notify the \wserver about changes in the ICN cache.

\wserver component is implemented as the Node.js application. It implements the generic web application model as a set of HTML templates with empty {\em div}s, which are filled with script outputs. The HTML template engine is implemented using EJS\footnote{EJS: http://www.embeddedjs.com}. The generic web application model enables new content generation for the web users by providing them with an application specific HTML form. \wserver validates the filled form and sends it to \cgenerator using Redis queue. To obtain a good user experience, \wserver provides real-time content updates by means of WebSocket Protocol~\cite{rfc6455} that is implemented using socket.io\footnote{Socket.IO: http://socket.io} library. Finally, \wserver provides also ``app store'' functionality by allowing web users to upload their own native applications via an HTML form. 

\cgenerator is the Java native application that: (1) reads a new content request from the message queue sent by \wserver, (2) performs application-specific encoding of the new message and (3) publishes it to the ICN cache.

%The \presenter consists of the set of PHP scripts that: 1) communicates with the Redis to learn which application's content is available and 2) arranges presentation of content generated by transformations in the web browser. To improve user experience, the framework provides real-time content updates by means of the WebSocket Protocol~\cite{rfc6455} that is implemented using the socket.io\footnote{https://github.com/rase-/socket.io-php-emitter} library.
%
%The \webhandler is a set of three PHP scripts. The \textit{presentation.php} script returns a detailed view about the particular message. The \textit{new.php} and \textit{generate.php} scripts are called in case of generating a response to the existing message, or a very new message. The \textit{new.php} returns the web form generated by appropriate transformations to the \presenter, so that it can be shown to the user. The \textit{generate.php} script acts as a form handler that processes data submitted in this web form, and sends them to the \cgenerator for the message creation. To enable access to the whole content associated with the particular message in the Redis, all these scripts are executed with the Redis message key as the URL parameter. 
%
%The \cgenerator is the native opportunistic router application that: (1) reads data submitted by the \webhandler, (2) using
%the opportunistic router API, transforms them into an appropriate message and (3) publishes the resulting message in the opportunistic network.

\ifsubmission
\else
\begin{figure}[htb!]
  \begin{center}
    \includegraphics[width=0.85\columnwidth]{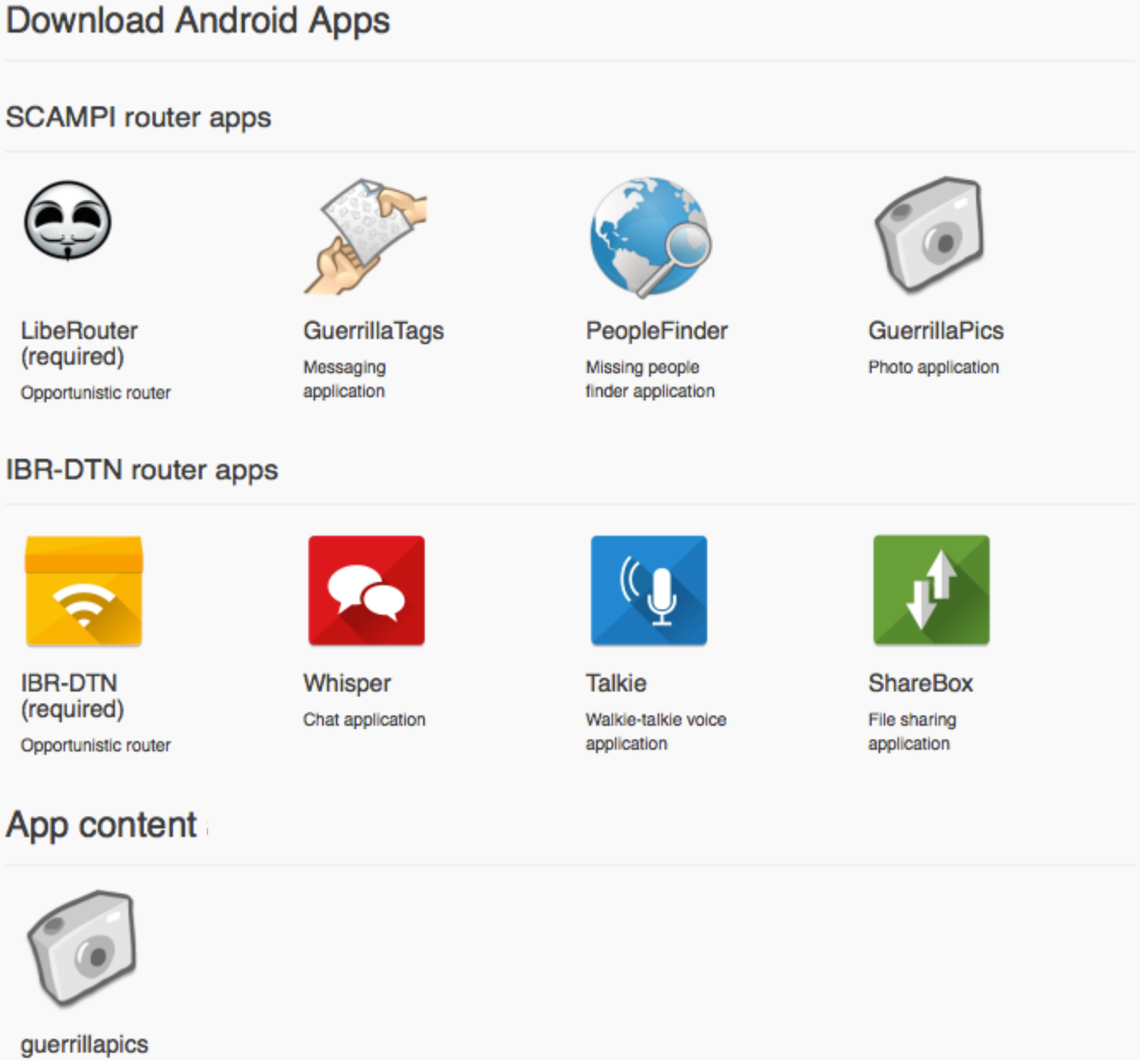}
    \caption{Screenshot of the framework presenting native opportunistic applications that can be downloaded on a device and applications having content stored on it.}
    \label{fig.webportal-screenshot}
  \end{center}
\end{figure}
\fi

\ifsubmission
\else
In our implementation there are no constraints on data types that can be attached to the message as part of meta data. The only requirement is that if an attached item is a Java serialized class, such a message must also carry a ``.class", or JAR file implementing the class. In such cases, the \ctracker dynamically loads attached data, transforms it to JSON\footnote{json-io: https://github.com/jdereg/json-io}, and stores it in the shared memory.
\fi

\descr{Security components details.} Recall from the Section~\ref{sec:security} that our threat model calls for implementation of the file system level isolation and the system call filtering. The framework realizes the file system level isolation by: (1) creating a temporal directory and inserting into it all data carried inside the message and (2) isolating other file system resources from the message content (by means of \chroot call). We realize the system call filtering using \seccomp~\cite{seccomp-bpf}. It allows to whitelist a subset of system calls that a transformation has a permission to execute~\cite{seccomp-bpf-kernel}. Invoking a system call that had not been whitelisted causes termination of the transformation process. Since the transformation should be given access to all meta data carried by the message, our \seccomp profile allows the execution of only system calls that open and operate on files. Whole functionality of \sandbox is implemented in C.
%\ifsubmission
%\else
%For system administrators concerned with \chroot vulnerabilities, it is trivial to further enhance sandbox design by replacing \chroot with \pivotroot \footnote{http://linux.die.net/man/2/pivot\_root} system call~\cite{thomas-nyman-thesis}. Unlike \chroot, \pivotroot confines the process into a completely separate root file system, thus even if the process manages to escalate privileges and theoretically break away from the file system isolation, it cannot access file system resources outside its directory, as these resources do not exist for it. Further security improvements can be made by defining mount and network namespaces for the sandbox and running transformations inside Linux containers\footnote{https://linuxcontainers.org}. However, we believe that for our use case it is an overkill.
%\fi

%\footnote{http://man7.org/linux/man-pages/man2/chroot.2.html} 
%\footnote{http://man7.org/linux/man-pages/man2/mount.2.html}
%Finally, in order to meet security requirements set in the section~\ref{sec:security}, the transformation is executed inside the separate application (developed in C) that acts as the sandbox.

Browser side verification of message authenticity requires availability of a public key in the browser via HTML5 local storage or access to device persistent storage (e.g., the File API~\cite{file-api}). Both of these functionalities are supported by all modern browsers. The authenticity verification process is realized fully inside the JavaScript code via Web Cryptography API~\cite{web-crypto-api}. To minimize trust put on the framework for verifying message authenticity, it is preferable for the user to use browser-side option of verifying authenticity. But the framework-side alternative (recall from Section~\ref{sec:security}) is the more realistic option due to poor support of Web Cryptography API among the modern browsers~\cite{web-crypto-support}.

\descr{Web application transformation.} The web application transformations are implemented in Python, and must follow basic transformation design guidelines described in Section~\ref{sec:conceptual-model}. Furthermore, to enable access to all meta data related to a particular message, the Redis key pointing to a location of item meta data in the shared memory is used as the transformation command line argument. In order to prevent the transformation from crashing caused by making a call to the library that is not available in the framework, the framework implements the Python module that checks for presence of a non-standard Python library. Consequently, if the non-standard library is not available, the transformation can make a fallback to a different call. Aiming for simplification of developer task to provide a good user experience, the framework gives the transformation access to Bootstrap\footnote{Bootstrap: http://getbootstrap.com} library, so that the transformation can generate a sophisticated HTML presentation of content, while content itself does not need to carry additional CSS style files as meta data.

\descr{Framework performance evaluation.} We evaluated our framework implementation by measuring execution time taken by each component to process a single message. We performed experiments on Macbook Pro (2.66 GHz Intel Core i7 CPU with 8GB of RAM) running OS X 10.10.1 and repeated the experiment 30 times to obtain statistical significance. 

\begin{figure}[htb!]
  \begin{center}
\ifsubmission
    \includegraphics[clip,trim=2cm 0.5cm 1cm 2cm,width=0.85\columnwidth]{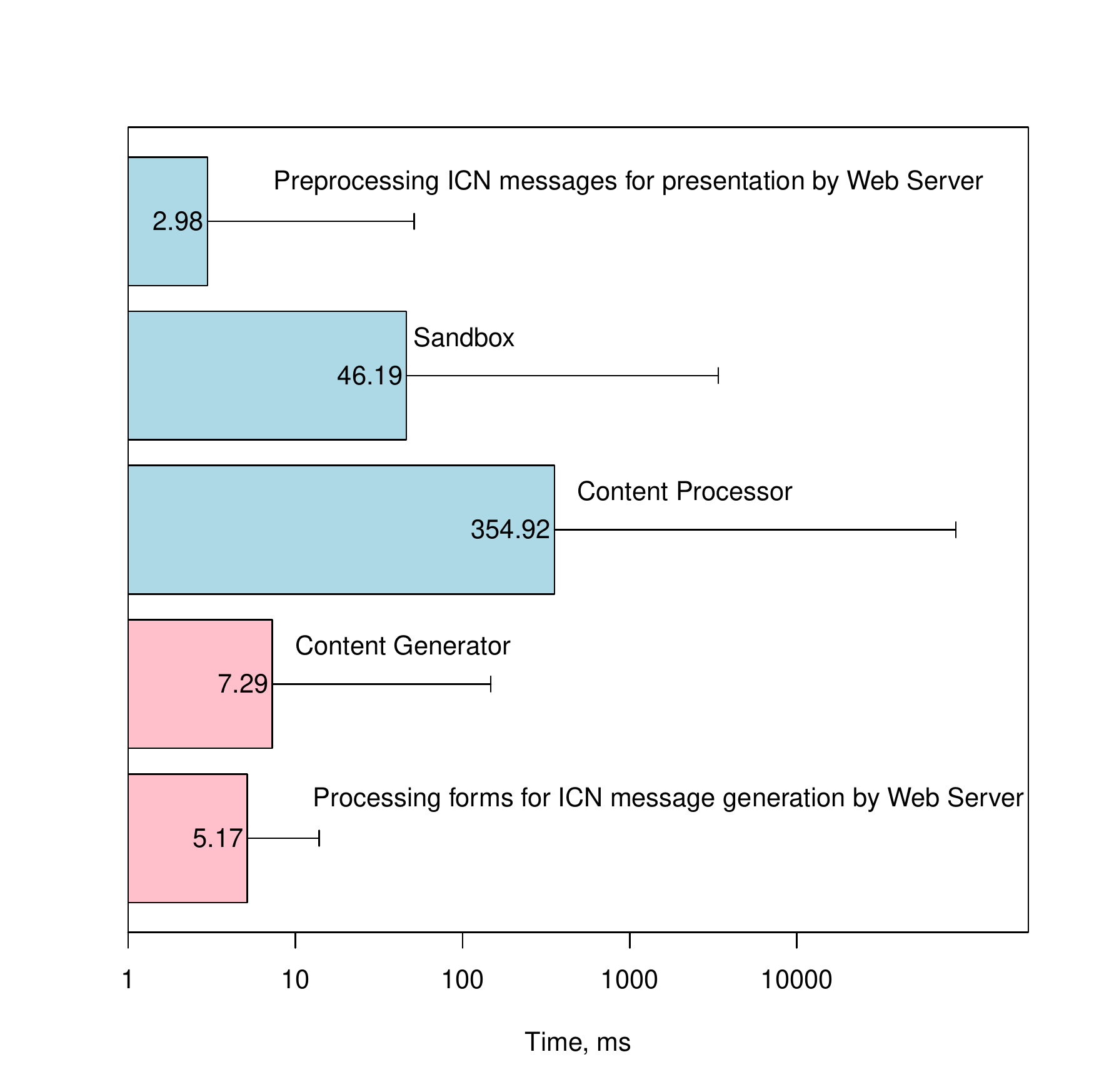}
\else
    \includegraphics[clip,trim=2cm 0.5cm 1cm 2cm,width=0.85\columnwidth]{processing.pdf}
\fi
    \caption{Mean execution time (measured in milliseconds) taken by each framework component (logarithmic scale).}
    \label{fig.perf-implementation}
  \end{center}
\end{figure}

\ctracker and \sandbox are the most resource consuming components (Fig.~\ref{fig.perf-implementation}). This is an expected result, as both of them perform higher number of I/O operations with the hard disk in comparison to other components. In overall, the average execution time on discovering new message in the cache is about $500$ms, while generation of a new message takes about $15$ms.

%\vspace{-0.5cm}
\section{System integration}
\label{sec:router-integration}

%% Teemu: This doesn't quite capture the idea.
%The framework resides in the throwbox and does not have a content dissemination functionality.
%
%Thus, to be usable it needs to be integrated with a third party system that provides interconnectivity between throwboxes.
%
%In practice, the framework can be integrated with any system which supports shipping meta data inside content.
%
%Such an integration requires implementing the ICN communicator interface (i.e., monitoring of ICN cache changes and new content publishing).
%
%Since our work focuses on disconnected environments, we have integrated our framework with two widely used delay tolerant network
%(DTN) systems, namely \scampi and \ibr.
 
The framework is abstracted from the details of the underlying caching mechanism through \icomm interface.
In the simplest case, the operation are purely local filesystem reads and writes, and no dissemination happens beyond
\liberouter device.
To enable dissemination beyond the local device, the framework can be integrated with some ICN platform, which
implements content spreading between nearby devices.
In practice, the framework can be integrated with any system which supports shipping meta data inside content.
To demonstrate this concept in practice, we have implemented integration with two disconnected networking solutions,
\scampi and \ibr.
In both cases \icomm reads messages directly from the platform cache, and writes messages through
the respective platform API.

%\ifsubmission
%\else
%In addition, we also offer an opportunity to use the framework that is not coupled to any distribution of opportunistic router (so called \nil router version).
%\fi

\descr{\scampi.} Although \scampi is designed to be the delay tolerant networking (DTN) platform, it meets all requirements of the ICN platform. As such, it provides: (1) data cache, (2) publish-subscribe functionality and (3) the network layer implementations capable of delivering messages based on store-carry-forward networking. It is implemented as the middleware that runs on any platform with Java support, thus it allows to build an efficient ICN network with mobile devices acting as ICN nodes.
%
%It discovers and opens links between nearby peers using Wi-Fi and Bluetooth. 
%It uses Wi-Fi and Bluetooth as the physical layer technology to route messages between nodes. The underlying design follow the DTNRG architecture and protocols~\cite{rfc4838,bundle-spec,dtn-tcp-cl}.
\ifsubmission
Our tech report~\cite{dtn-legacy-app-framework-full} contains more details on \scampi platform.
\else
%It uses Wi-Fi and Bluetooth as the physical layer technology to route messages between nodes. The underlying design follow the DTNRG architecture and protocols~\cite{rfc4838,bundle-spec,dtn-tcp-cl}.
It also includes a number of extensions to support geographically constrained content distribution~\cite{floating-content}, and content search mechanisms. Among existing opportunistic network routing protocols, only Epidemic~\cite{epidemic} and static routing are supported. Namespaced metadata key-value pairs can be attached to the messages to provide application hints to the underlying networking layer.
\fi

Applications can be developed to use the communication services provided by \scampi. These applications can either be native (Java) applications, or HTML5 web applications. 
\ifsubmission
\else
The applications can also be distributed by the middleware without requiring a centralized app store. However, to run the applications, the middleware and the application itself must be running on the device.
\fi

Implementation of \icomm for \scampi requires: (1) developing cache monitor as a directory tracker (\scampi implements cache as a set of file system directories) and (2) using \scampi {\em AppLib} library to publish new content to ICN.

%: 1) allowing the framework to access \scampi's cache and 2) implementing new content creation functionality in the \scampi compliant format. The former one is accomplished by granting the framework access rights to the \scampi's cache location, as \scampi implements cache as a set of file system directories. The latter one is achieved by using \scampi's {\em AppLib} library that allows to publish new content to the \scampi's ICN node.

\descr{\ibr.} \ibr is the DTN platform, similarly to \scampi working as a middleware. Unlike \scampi, it does not provide publish-subscribe functionality, but it is also the data dissemination system with cache functionality, thus fitting into our scope of disconnected environments. 
%It runs on any Linux based operating system as well as on Android and BeagleBone\footnote{http://beagleboard.org/bone}. It discovers and establishes communication links with nearby devices using Wi-Fi, Wi-Fi Direct~\cite{wifi-direct} and Wireless Personal Access Network(IEEE 802.15.4)~\cite{lr-wpan} technologies. Like \scampi, the \ibr's design follows the basic DTNRG architecture~\cite{rfc4838,bundle-spec,dtn-tcp-cl}. 
\ifsubmission
More details on \ibr platform are available in our tech report~\cite{dtn-legacy-app-framework-full}.
\else
It runs on any Linux based operating system as well as on Android and BeagleBone\footnote{http://beagleboard.org/bone}. It discovers and establishes communication links with nearby devices using Wi-Fi, Wi-Fi Direct~\cite{wifi-direct} and Wireless Personal Access Network(IEEE 802.15.4)~\cite{lr-wpan} technologies. Like \scampi, the \ibr's design follows the basic DTNRG architecture~\cite{rfc4838,bundle-spec,dtn-tcp-cl}.
In addition to \scampi, it implements also convergence layers for other underlying network technologies (i.e., UDP over IP, IEEE 802.15.4 LoWPAN). Unlike \scampi, \ibr does not offer any application level extensions (e.g., content search mechanisms), however, in addition to the Epidemic routing, it provides other opportunistic network routing protocols, namely PRoPHET~\cite{dtn-prophet-id} and Direct Delivery routing. 
\fi

Applications can be developed for \ibr communication service with Java and C++ client libraries currently available to developers. 
\ifsubmission
\else
For other programming languages, developers must write their own client libraries to communicate over TCP with middleware services.
\fi

The process of \ibr integration into the framework is identical to the \scampi integration. The only difference is that developers must replace {\em AppLib} with: 1) {\em ibrdtnlib} as a Java library, 2) {\em libapi} for C++ development and 3) {\em ibrdtn-api} for Android.

%\ifsubmission
%\else
%\descr{\nil router.}
%In this mode, the framework cannot take any advantage of libraries provided by opportunistic network platform. Framework access to the router DB is achieved in identically as for \scampi and \ibr. However, creation of new messages by the framework requires it to: 1) manually build a message (according to the Bundle Protocol~\cite{bundle-spec})  and 2) copy the created message to the router DB.
%\fi

%\subsection{Application integration}
%In all applications integrated into the framework, we have extended their message format by adding: application presenter, message summary, new and presentation transformations, application icon, and application name as meta data. In addition for {\em PeopleFinder}, {\em Whisper} and {\em Talkie}, we have also included the response transformation. 

% \descr{Source code.} The source code of the framework implementation together with the enhanced applications will be made available with the final version of the paper.

\section{Sample Web Applications}
\label{sec.applications}

%% Teemu: These scripts process single messages, independently. Do we have
%% a concept of aggregate processing? For example, to generate the presentation
%% for GuerrillaTags, we want to create a list of all unique tags contained within
%% all the tags messages. => per-service scripts

%We have used the framework to develop enhancements to six existing open-source opportunistic applications working for Android.
%
To build the web version of the existing native application, an application developer must implement transformations in Python needed in the application. 
Among all defined transformations, only the {\em message summary} transformation is necessary, as it allows the framework to always interpret the content of the particular message independent of presence of other messages belonging to the web application (but content generation and application level logic is not available then). 
Developers attach transformations to messages using third party software libraries (recall from the Section~\ref{sec:router-integration}).

To facilitate development process, we have also implemented a simple transformation testing environment. 
It allows the application developer to test correctness of his implementation by: (1) executing the transformation in it and (2) verifying that the HTML view generated by the transformation is coherent with the desired application design.

Now we present description of our six web applications.
From the available \scampi applications we have extended {\em GuerrillaPics}, {\em GuerrillaTags} and {\em PeopleFinder}, while for the \ibr router we enhanced {\em Whisper}, {\em Talkie}, and {\em ShareBox}. As {\em PeopleFinder} application is the most sophisticated one among these applications, we describe it in the Section 6.1. The other applications are described in the Section 6.2.

%
%All of these applications present different types of interactions, but can be enhanced by our framework.
\subsection{PeopleFinder application}
\scampi's version of {\em PeopleFinder} is an adaptation of Google Person Finder\footnote{https://google.org/personfinder/global/home.html} application into disconnected environments. The Google's version of the application has proven its usability in various disastrous scenarios starting from 2010 Haiti earthquake.
The application allows missing persons records to be generated, and notes to be attached to those
records.
Each message generated by the application contains the record and a set of all notes
related to the record known by the sender.
The application summary view is generated from the person records.
There may be multiple records for the same person, in which case each one is listed
separately.
%
%In addition to listing the records, the generated view also includes JavaScript based
%search functionality, that allows the users to filter the view based on search terms.
%
Notes attached to a person are displayed in the detailed view, which is generated
by the {\em message presentation} transformation.
Adding a new note for an existing record is done through the {\em reply} transformation,
which gets the original record (and notes) as a parameter.
This transformation appends the new note to the existing ones, and generates a new aggregate
message with the record and all known notes.
A brand new missing person record is generated through the {\em new} transformation.
Figure~\ref{subfig:people-finder-compare} shows the native PeopleFinder, and its framework version.

\subsection{Other applications}

\begin{figure*}[htb]
\centering
	\begin{subfigure}[b]{\columnwidth}
		\centering
\ifsubmission 	
		\includegraphics[width=0.75\linewidth]{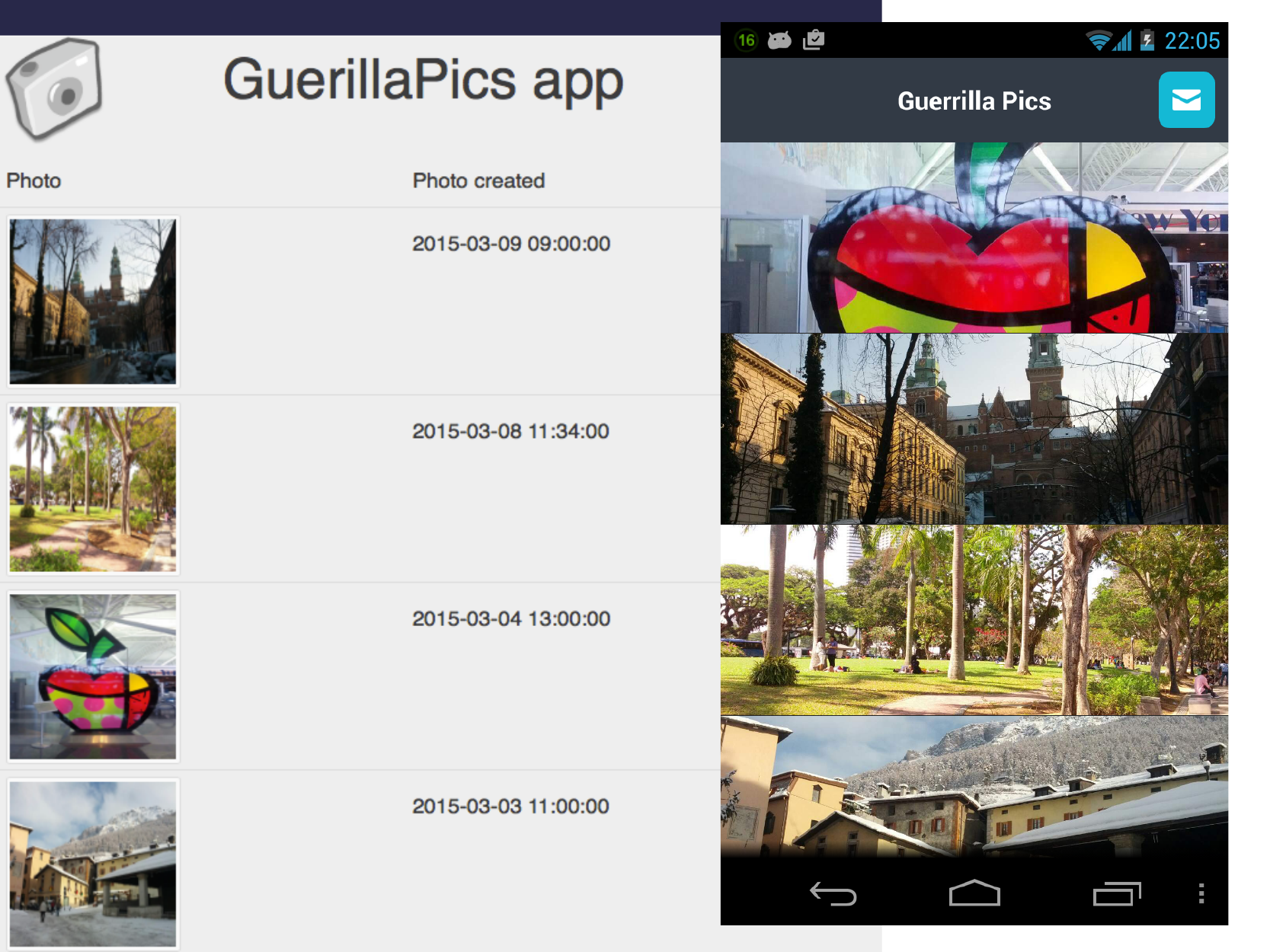}
\else
		\includegraphics[width=0.75\linewidth]{guerrilla-pics-comparison-2.pdf}
\fi
 		\caption{GuerrillaPics}
                 \label{subfig:guerrilla-pics-compare}
        \end{subfigure}
	\qquad
	\begin{subfigure}[b]{\columnwidth}
		\centering
\ifsubmission 	
		\includegraphics[width=0.75\textwidth]{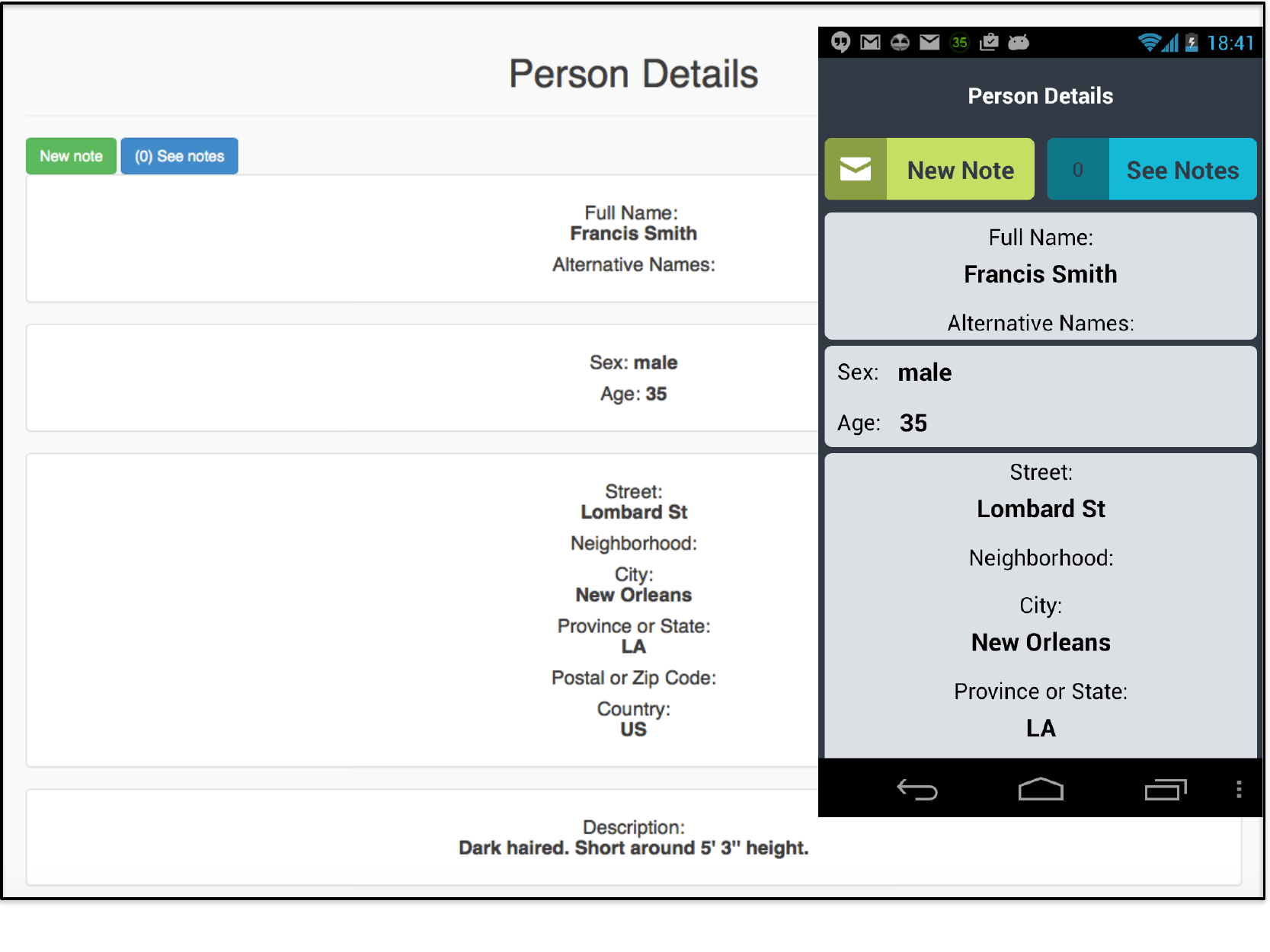}
\else
		\includegraphics[width=0.75\textwidth]{people-finder-comparison-2.pdf}
\fi
 		\caption{PeopleFinder}
                 \label{subfig:people-finder-compare}
         \end{subfigure}
    \caption{Comparison of native and web-based user interface for: a) GuerrillaPics, b) PeopleFinder apps.}
     \label{fig.app-compare}
 \end{figure*}

{\em GuerrillaPics} is the \scampi photo sharing application.
Since its messages have no dependencies between them (e.g., there are no replies or groupings), this is the simplest use-case for the framework.
First, an {\em application presenter} transformation is used to generate a grid of 
photo thumbnails.
The transformation calls the {\em message summary} for each message to generate a thumbnail and to get the creation timestamp.
These summary elements are then listed in a grid based on time ordering and displayed to the users through the framework.
If a user selects one of the summaries by clicking the thumbnail, the {\em message presentation} transformation is used to generate a full resolution view.
Finally, the application has also the ability to share a new photo through the framework by means of the {\em new} transformation.
Figure~\ref{subfig:guerrilla-pics-compare} illustrates comparison between native GuerrillaPics, and its framework version.

{\em GuerrillaTags} is a message board application where messages are not
independent as the photos in the previous case, but rather exist in
the context of a topic ({\em tag}).
In this case, the summary view is composed from all the unique tags contained
in all the messages that belong to the application.
This is done by the {\em application presenter} transformation, which: 1) applies the {\em message
summary} transformation to all the application messages and 2) filters out duplicates from the produced 
list of tags to the list of available topics (each unique topic is a summary item).
Each topic has its own state entry, which contains the set of messages with the common tag.
The presentation view is per-topic, and lists all messages belonging to the
given topic.
It is generated by another presenter transformation by applying the {\em message
presentation} transformation on all the relevant messages (as determined from the previously
saved topic state), and sorting the resulting list by creation timestamp.
This demonstrates how complex application level logic can use the application
transformations to generate complex presentation views.
The application also allows to post messages via the framework by using the {\em new} transformation.

All \ibr applications use unicast communication model. Since the framework currently does not support this communication model (recall Section~\ref{sec:security}), we have modified these applications to use the group communication model.

{\em Whisper} is an opportunistic network chat application in which exchange text messages within a group of devices.
Thus, the messaging model and application logic are identical to the GuerrillaTags application with the group of devices identifier (group ID) playing the role of the GuerrillaTags's topic. 
Whisper includes additionally the functionality of responding to the current chat discussion through the framework via the {\em reply} transformation.

{\em Talkie} is an opportunistic network walkie-talkie voice application in which users share voice messages with a group of devices. 
The messaging model and transformations logic for the message presentation are identical to Whisper's use case (only chat message is replaced by voice).
To generate a new message (or respond to existing conversation), Talkie uses {\em new} and {\em reply} transformations.
They use the Media Capture API~\cite{media-api} for recording the audio messages.\footnote{Not currently supported by Safari and Internet Explorer.}
%%
%The application summary view is built from all unique device groups contained in all the app messages. 
%%
%This view is generated by the {\em application presenter} transformation. 
%%
%It executes the {\em message summary} transformation, which return group ID for each message. 
%%
%After that the {\em application presenter} transformation filters out duplicate group IDs. 
%%
%The presentation view is per group ID and contains the list of all voice messages belonging to the given group ID. 
%%
%All these voice messages have the ability to be played out via the {\em message presentation} transformation. 
%%
%Finally, Talkie allows users to send new voice messages, or respond within a group of devices via the framework using the {\em new} and {\em response} transformations. 
%%
%These transformations make use of the JavaScript API for media capture~\cite{media-api} to give web browser access to device's microphone for voice recording. 
%%
%Unfortunately the media capture API is not currently supported by Safari and Internet Explorer.
 %As the application interactions are identical to Whisper (only text content is replaced by voice), the logic of message transformations is similar to the Whisper's use case. The summary view

{\em ShareBox} is an opportunistic network file sharing application in which users exchange pictures or other files within the given group of devices. 
Since its messages have no dependencies between them, the messaging model and transformations logic is similar to the GuerrillaPics app. 
%
%%
%The messages are self-contained, stateless, and do not have any dependencies between themselves. 
%%
The application summary view %(generated by the {\em application presenter} transformation) 
shows a table of message summary views for all Sharebox messages. 
The message summary view %is created by the {\em message summary} transformation which
shows the size of the message together with the device identifier of its sender and the timestamp. 
Finally, if the user selects a particular message, the {\em message presentation} transformation shows the actual files carried inside the  message.

%% Do we need this? Does it add something fundamentally different to the other ones?
%% Do we want to have it for completeness?
%{\em nearbyPeople} is an opportunistic social networking application, which lets
%users publish their profile into the opportunistic network.
%

%
%-- PeopleFinder??
%
%-- nearbyPeople??

\section{Validation}
\label{sec.validation}
% Simulation ideas
% -- Access of legacy nodes to contents (fraction legacy compared to
%    fraction native)

Our  framework offers, in principle, content
access to web users (in addition to native app users).  To achieve this,
the framework requires code to be shipped along with message state
updates, which incurs overhead.  In the following, we first evaluate
the overhead and its impact.  We then turn
our attention to how many web nodes could be \textit{reached} by
content if those nodes choose to look at messages.

A related question is if web users moving between access points would
also contribute to the connectivity of a disconnected ICN.  We have
shown that they can make a difference if the Liberouter nodes instrument
the web storage of mobile browsers for relaying messages \cite{dtn-legacy-routing}.

%We are interested in the following questions:
%\begin{itemize}
%    \item How does the overhead imposed by the framework affect the system
%        performance?
%    \item What is the impact of introducing more legacy nodes into the network?
%    \item How do legacy nodes participate in distributing messages and does it
%        provide benefits for the DTN nodes?
%\end{itemize}

\subsection{Overhead} 

To evaluate overhead, we measure the actual message sizes of
our implementation for three applications with sample contents.  The
results are shown in table \ref{tab.overhead}): When using a text
messaging application with tiny content, the message size grows almost
50-fold.  However, this is only due to the small size of the native
messages.  If the content size of the application messages increases,
the overhead becomes more reasonable.  For photo sharing, with small
photo size of 65--120\,KB, including the framework code adds just
5--10\% overhead.  For the most sophisticated application we looked
at, PeopleFinder, the overhead is roughly factor 15.

\begin{table}[tb!]
  \small
        \centering
	\begin{tabular}{|c|c|c|}
	\hline
	\textbf{Application} & \textbf{Native message size} & \textbf{Framework overhead} \\ \hline
	Text messaging & 350\,B & 16\,KB \\ \hline
	Photo sharing & 65--120\,KB & 7\,KB \\ \hline
        PeopleFinder & 2\,KB & 28\,KB \\ \hline
	\end{tabular}	
        \caption{Overhead introduced by the framework.}
        \label{tab.overhead}
\end{table}

Obviously, the code added via the framework is a function of the
complexity of the code required to interpret, render, and construct
messages: simpler applications will need less code.  The overhead is
obviously also a function of the content size so that more elaborate
content will cause, even if more complex code is needed, limited
overhead only.  One can argue that trends in web site complexity and
size\footnote{www.websiteoptimization.com/speed/tweak/average-web-page/}
show that increasing amounts of effort are put into conveying probably
roughly the same amount of content.  Thus, adding more sophisticated
interaction framework code for a better experience would just mirror
what is already done on the web.

For disconnected ICNs exploiting opportunistic encounters of mobile
nodes, the most important question is if and how the larger message
sizes affect message delivery performance.  To this end, we carried
out simulations using the ONE simulator \cite{theone} with two
different mobility models: 1) SPMBM: Shortest Path Map-Based Movement
between waypoints chosen from the Helsinki downtown map
(4.5$\times$3.4km$^2$) \cite{theone} for 50, 100 and 200 pedestrians
moving with speeds $v=U(0.5,1.5)$m/s without predefined points of
interest.  2) Same as 1 but additionally we introduce 10, 65 and 325
static access points.  3) SMOOTH: a simple way to model human
walks \cite{SMOOTH-trace-based-mobility} with the map from KAIST
scenario (10$\times$18km$^2$) and 50 nodes.  SMOOTH is a synthetic
model but it is based on real traces and captures most of the
properties of human mobility.
Our nodes communicate at a net bit rate of 2\,Mbit/s with a radio range of
50\,m.  The nodes use simple epidemic routing \cite{epidemic}.  We
choose a random node to generate a new message every 12\,s, 60\,s,
and 300\,s, referred to as high, medium, and low load, respectively.
Messages expire after 5400\,s.  The message sizes correspond to those
for native and framework-enhanced messages for the text and photo
sharing applications to pick two extremes.
We measure the fraction of nodes that obtain a copy of each message,
termed \textit{coverage}, and plot the average of 10 simulation runs,
each lasting for 12 hours.

\begin{figure}
  \begin{center}
\ifsubmission
  \includegraphics[clip,trim=0cm 0cm 0cm 0.3cm,width=6.3cm]{plots/results/spmbm/overhead/overhead.pdf}
  \includegraphics[clip,trim=0cm 0cm 0cm 0.3cm,width=6.3cm]{plots/results/spmbm-pics/overhead/overhead.pdf}
\else
  \includegraphics[clip,trim=0cm 0cm 0cm 0.3cm,width=6.3cm]{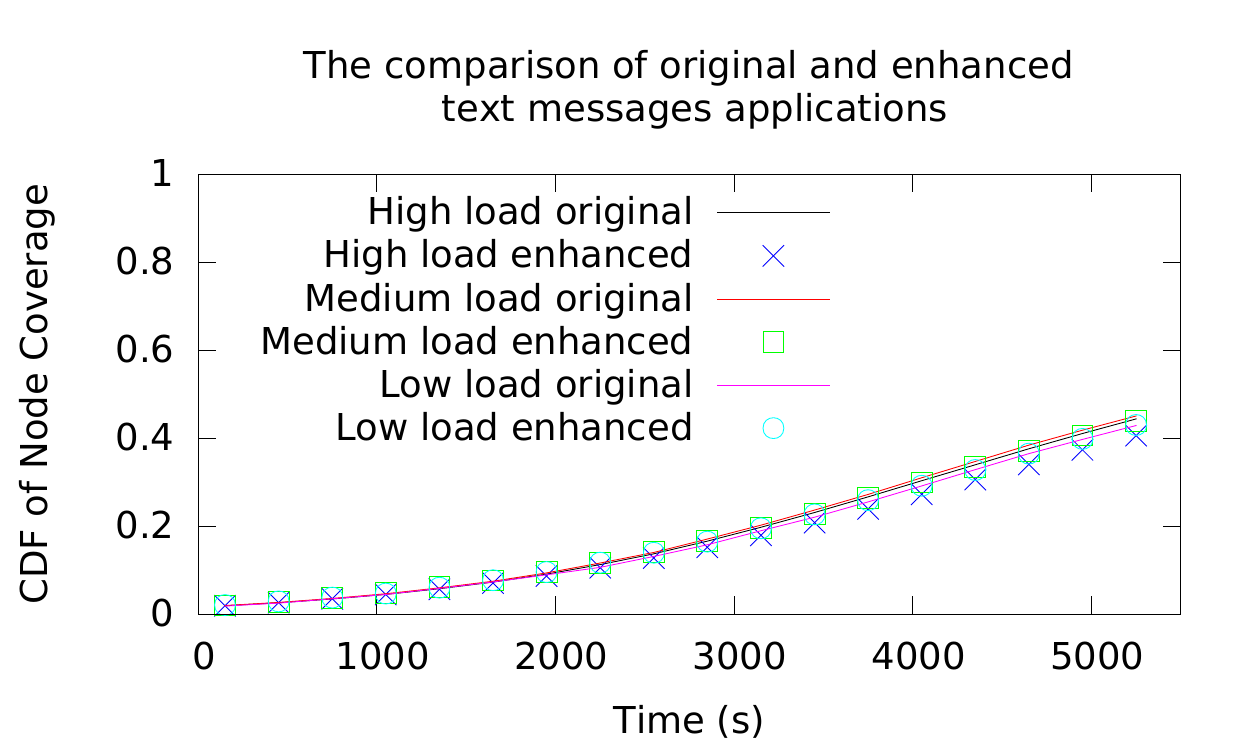}
  \includegraphics[clip,trim=0cm 0cm 0cm 0.3cm,width=6.3cm]{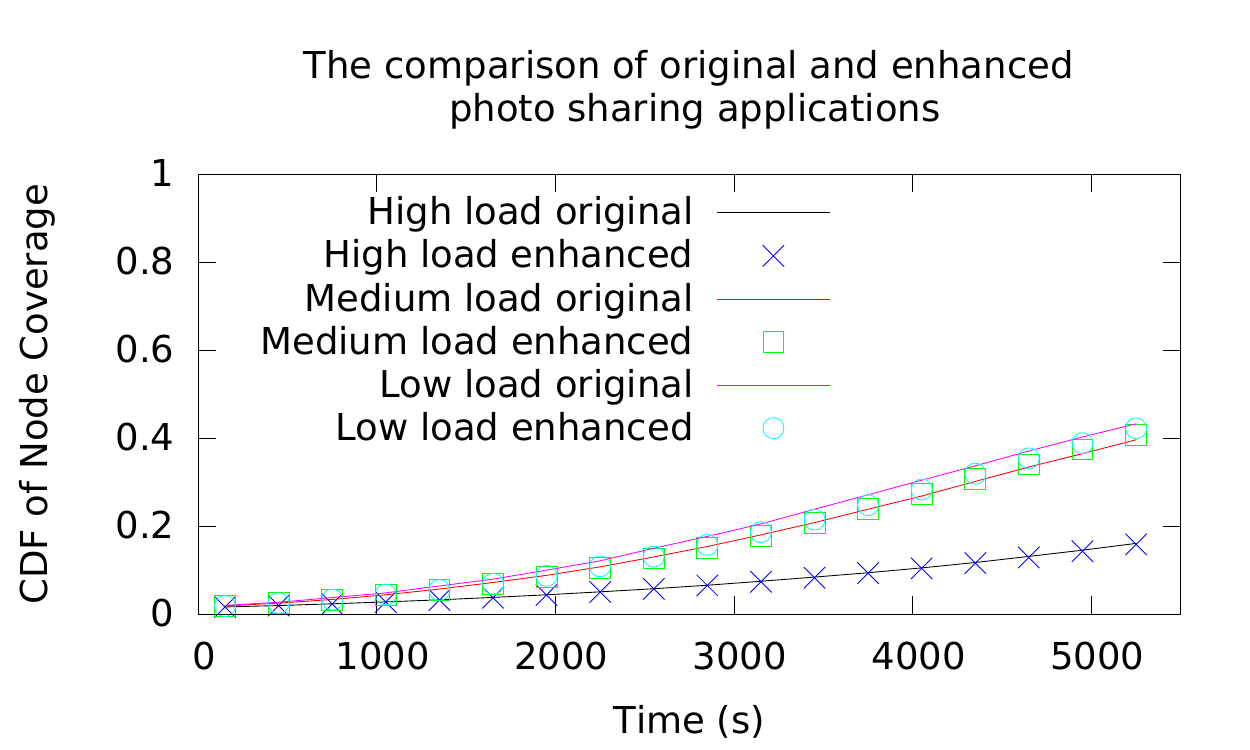}
\fi
  \caption{Impact of the overhead introduced by the framework (reflected in different message sizes) on the coverage
  for messaging (top) and photo sharing (bottom) with all three loads for the SPMBM model.}
  \vspace{-3mm}
  \label{fig.overhead}
  \end{center}
\end{figure}

As shown in Figure \ref{fig.overhead}, we find that the overhead of
the framework does not notably impact the performance results.  The
coverage remains the same both for using native application messaging
and for the framework-enhanced messaging for the text chat application
(top) as well as for photo sharing (bottom).  For the SMOOTH mobility model
we obtain similar results just with much lower
coverage rate due to sparser node distribution.
This indicates that in the realistic parameter range, the system is
not bottlenecked by the contact capacity, and therefore the added
messaging overhead does not negatively impact the message distribution.
%In the simulations,
%this can be explained by the messaging implementation that does not
%allow transmitting more than one message per simulator clock tick
%(which is 0.1s and yields 25\,KB per tick).  Yet, we obtain similar
%results for photo sharing (bottom), where the messages are larger than
%a clock tick and the sizes variable.

Further, a maximum message rate limit was also observed in past
experiments, which have shown that the per-message overhead of
protocol implementations appears to be more dominant in communication
performance than the per-byte overhead.  This was, for example, found
in a comparison of three different DTN bundle protocol implementations
\cite{dtn-impl-performance}, particularly for growing the payload size
from 10 bytes to 10\,KB and beyond.  Our own (not yet statistically
significant) experiments seem to confirm this.  While implementation
details play one important role here, there are also systematic aspects
to consider: nodes that meet need to exchange vectors which messages
they have, decide which ones to replicate, and then perform a
forwarding process for each message, which causes per-message
overhead.  Researchers also found that neighbor discovering and pairing
with peers is expensive and takes easily tens of
seconds \cite{experimenting-opp-net} while a 20\,KB data transfer
takes only 160\,ms even assuming just 1\,Mbit/s data rate.  We
therefore argue that the framework overhead is not of substantial
importance in practice.

\begin{figure}
  \begin{center}
\ifsubmission
    \includegraphics[clip,trim=0cm 0cm 0cm 0.3cm,width=6.3cm]{plots/results/half_ap_half_dtn/legacy/legacy.pdf}
    \includegraphics[clip,trim=0cm 0cm 0cm 0.3cm,width=6.3cm]{plots/results/half_ap_pics/legacy/legacy.pdf}
\else
    \includegraphics[clip,trim=0cm 0cm 0cm 0.3cm,width=6.3cm]{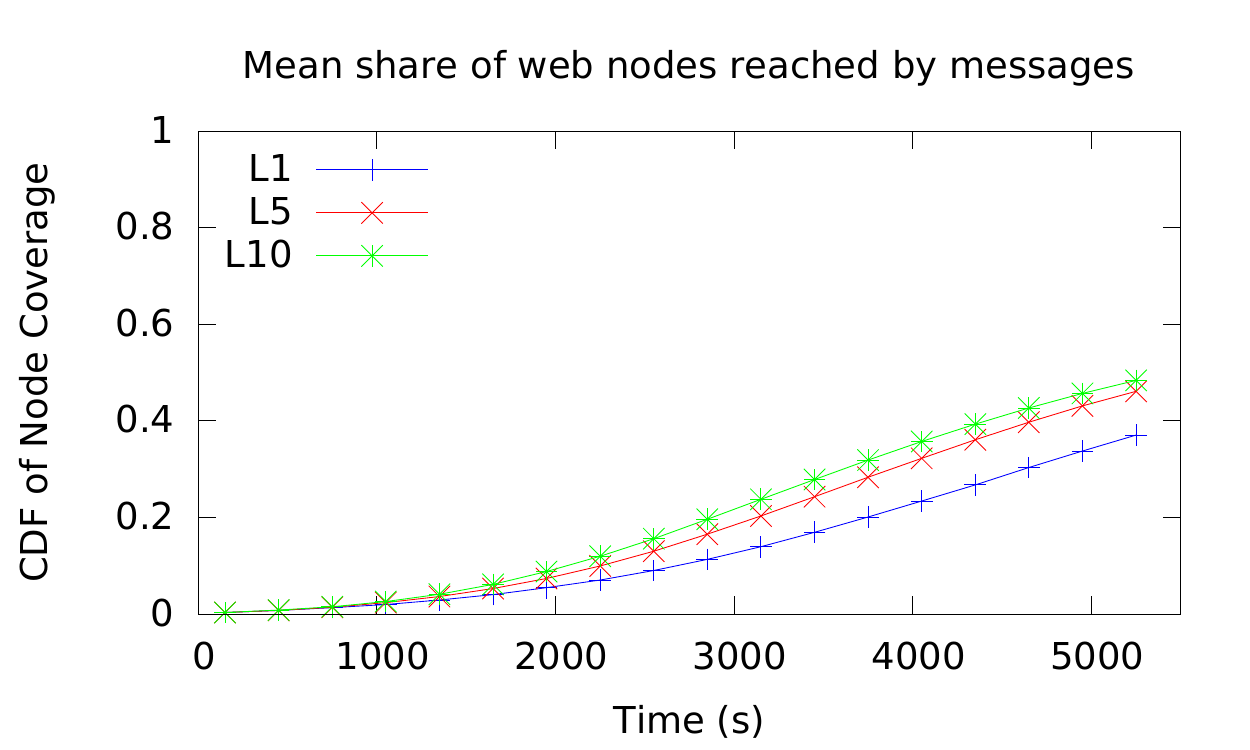}
    \includegraphics[clip,trim=0cm 0cm 0cm 0.3cm,width=6.3cm]{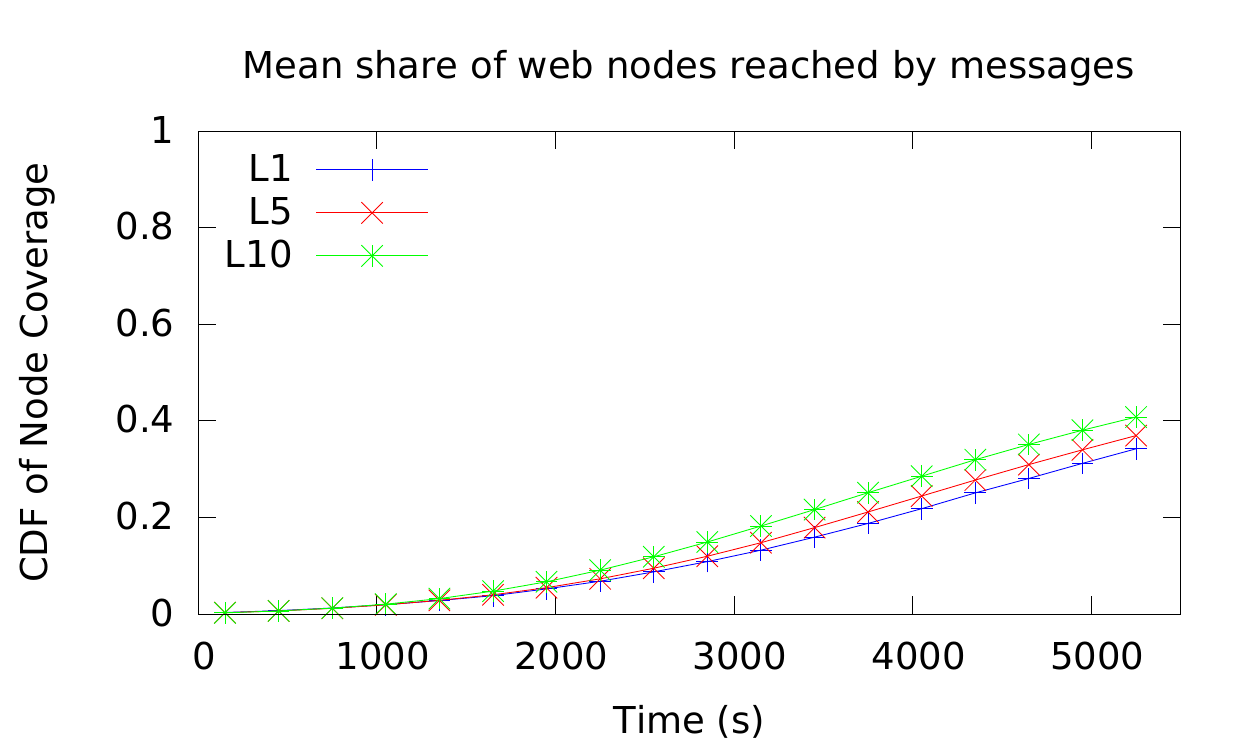}
\fi
    \caption{Coverage achieved for 50, 250 and 500 web nodes
    for the SPMBM model with 50 native nodes of which 25 are moving access points with medium load for
    text messaging (top) and photo sharing (bottom).}
    \vspace{-3mm}
     \label{fig.coverage}
  \end{center}
\end{figure}

\subsection{Content Reach}

The previous subsection suggests that the overhead introduced by our
framework won't degrade performance for native nodes.  But how well does
the framework allow reaching out to web nodes?  We conduct further
simulations to answer this question, using largely the same setup as
above.  In addition, we introduce two further classes of nodes:
\textit{access point nodes} that, besides running the ICN middleware
protocols, also serve as WLAN access points and run the server side of
the interaction framework (cf. Figure \ref{fig.system}).  And \textit{web
nodes} that only interact with these access point nodes, but neither with
each other nor with regular DTN nodes.  We choose the number of web
nodes to be equal (L1), five-fold (L5), or ten-fold (L10) the number of
DTN nodes.

Obviously, how many web nodes we can reach will depend on the movement
patterns of those nodes and where the access point nodes are located,
and how they move.  However, our simulation results shown in
Figure \ref{fig.coverage} hint that we can notably increase the
visibility of content generated by native applications; without the
web framework this content would not be accessible by web nodes.  For
the SPMBM scenario, we find that the content coverage may come close
to that of the native nodes and reach close to 40\% of the web
nodes. Additionally deploying 10 static access points brings it up to
50\%.
Comparing this to
Figure \ref{fig.overhead}, content availability is roughly equal for
native and web nodes.  Note that, the fraction of web nodes
reached gets bigger as their number increases from L1 to L5 to L10,
so the absolute number of additional nodes reached grows even more.
We obtain similar findings for high (up to 40\%) and low loads (up to
50\%) for both text and photo applications.  The SMOOTH scenario
yields a qualitatively similar picture, again at much lower
performance.  Results of our simulations with 100 and 200 native nodes
for all scenarios are also in line with these findings.

\section{Related Work}
\label{sec.related}
Our framework borrows concepts from different fields of related work
to create a unique combination.  Most important is the concept of 
%
%We begin with research work targeted at making the core network
%programmable. Motivated to accelerate of Internet evolution,
embedding programs into messages and executing them in network nodes,
discussed in the past as active networking \cite{active-networks} and
mobile code.  Lee et al.~\cite{Lee11Netserv} present a a node
architecture allowing to deploy in-network services in a next
generation Internet.  Its main contribution is the concept of making
the core network become a distributed service execution environment.
It also describes an architecture for extensible router allowing for
implementing new router features. Similar concepts of extensible
router architecture can be also seen in works of Router
Plugins~\cite{Decasper98routerplugins},
LARA++~\cite{conf/iscc/SchmidFSS01}, PromethOS~\cite{Kel02b},
Pronto~\cite{pronto} and SARA~\cite{conf/qofis/BagnuloACS02}.
SOFTNET~\cite{softnet}, PLAN~\cite{PLAN}, Bowman and
CANEs~\cite{bowmanAndCanes} are examples of active networking systems
that assume network packets to contain programs, which are used to
manage network nodes.  All these systems concentrate on executing code
inside the network in order to improve network capabilities, while our
solution takes advantage of transformation execution to enable content
access to legacy users.  Moreover, active networks focus on
individual (small) packets, thus limiting the amount of code that 
can be carried, while our message-based system is not limited by 
MTU size.

Baldi et al. \cite{conf/icse/BaldiP98} and
Thorn \cite{Thorn:1997:PLM:262009.262010} study the applicability of
programming languages in mobile code.  Ghezzi et
al.~\cite{conf/ma/GhezziV97} describe architectures of mobile code
applications. Our framework choose one specific programming language
and a specific application design tailored to the purpose of read
and write access to opportunistic message contents.

Security aspects of mobile code are covered by Arden et
al.~\cite{Arden:2012:SMC:2310656.2310677}.  He introduces a new
architecture for secure mobile code that developers can use, publish
and share mobile code securely across trusted domains.  Older work
presenting security aspects of mobile code are Rubin et
al.~\cite{DBLP:journals/internet/RubinG98},
Zachary~\cite{journals/internet/Zachary03} and
Brooks~\cite{DBLP:journals/internet/Brooks04}.  Kosta et
al.~\cite{DBLP:conf/infocom/KostaAHMZ12} present the concept of using
mobile code for improving code execution on mobile devices by
offloading part of code execution to the cloud.  Similar work by
Simanta et al.~\cite{DBLP:conf/wicsa/SimantaLMHS12} describe an
architecture for offloading mobile code execution in hostile network
environments.  Unlike these works, our system uses mobile code as a
tool for message content presentation and our main concern is offering
an isolated execution environment for the application code so that
the code does not harm the node running it (which is quite similar to
the concerns of protecting routers in Active Networks).

Another set of related work concerns document presentation techniques.
MINOS~\cite{Christodoulakis:1986:MDP:9760.9764} was an early system
allowing for presentation of multimedia content embedded in a
document. Boguraev et al.~\cite{Boguraev:1998:DDP:279044.279160}
presents usage of linguistic analysis tools for generation of text
document description and its visualization. Our solution
is not limited to content presentation, but offers a full-fledged
interaction to the web user.

%Unlike these works, our solution
%mostly relies on transformations for message presentation, and methods
%of inferring message presentation are fairly simple.

Douceur et al.~\cite{Douceur:2008:LLC:1855741.1855765} shows an
alternative approach for using native applications in the web browser
by the legacy device users.  This system requires web servers to have
an (application-specific) gateway installed that translates a native
app into a web app. On the other hand, our framework is more flexible, as it carries
all transformation code inside the messages themselves so that no
node requires prior knowledge of specific applications.

\vspace{-3mm}
\section{Conclusion}
\label{sec.conclusion}

In this paper, we have presented an ICN based system design for
enabling modern web applications (i.e., interactive applications
that run in a web browser context) to be developed for and deployed
in disconnected networks.
Our design is based on the idea of distributing small, self-contained
pieces of application content directly in the network, rather than sending
it to a centralized database, and bundling the presentation and interaction
logic as code together with the message.
This approach enables browser-based interactions in
scenarios without a well-connected infrastructure network.

Beyond the design of our system, we showed and evaluated the practicality
of our approach through a prototype implementation on our \liberouter platform.
We showed generality of the approach by running our framework on
two different disconnected networking platforms, \scampi and \ibr.
Further, we demonstrated how to apply our design to multiple native
applications to produce an interactive web versions of them.
Finally, we showed through simulations that the overhead imposed by our
design is low enough that it does not have a significant negative impact
on the message dissemination by the underlying networking platform.

%In this paper, we have presented the concept of web browser based framework that provides
%legacy users the ability to take part in an opportunistic network with
%nothing more than a standard web browser.
%%
%The users can both see the content published by native applications and
%publish content themselves.
%%
%We showed through simulations that this significantly increases the reach
%of the content both for legacy and native users, with overheads that
%do not impose a significant penalty on message delivery.
%%

%We have described our implementation of our framework for two popular
%opportunistic networking platforms, and enhanced a total of six
%applications for those platforms.
%%
%This has a potential to significantly increase the attractiveness of
%both networks, as legacy users can access the content without having
%to install any extra software -- a common issue for opportunistic
%network deployments in the real world.

%We have implemented our system in the \liberouter, and enhanced three
%existing opportunistic applications to be supported by the \framework.
%Finally, we have also shown that by providing legacy  users content access
%and by using these devices to forward small messages, the content
%\textit{coverage} in the opportunistic network improves.

There are a number of open issues to be addressed in our future work.
This includes enhancing security and privacy functionality of the
framework by enabling access to encrypted content for web users and
allow the implementation of closed communication groups.  One special
case of closed groups will be embracing web users also for the
point-to-point (user-to-user) unicast communication. Finally, we are
exploring how to exploit web browser capabilities for establishing
direct browser-to-browser communication for web users without a
mediating entity (such as an access point).

%% Teemu: We don't have time to provide the full images here since
% the Edison stuff still needs some manual tweaks after flashing and
% the RPi images are out of date.
The framework and simulation source codes are available at \linebreak
http://www.netlab.tkk.fi/tutkimus/dtn/legacy-app-framework/.

%In this paper, we have explored the idea of making contents exchanged
%in opportunistic network accessible to users who do not run the
%corresponding software on their devices and leveraging their devices
%to assist in message spreading. Theory suggests a positive impact on
%overall system performance, which our simulations confirm for most of
%the---still limited---scenarios we explored.  We implemented our
%design in our \liberouter and carried out a set of initial
%experiments, showing that both web-based interaction and message
%carriage are feasible.  One open issue is how to further automate the
%instrumentation of legacy nodes, so that message forwarding could
%operate in the background.
%
%Besides a broader performance evaluation (also using traces if
%suitable ones become available), future directions include more
%comprehensive support for presentation and interaction code embedded
%in messages, adding security for \textit{read} and \textit{write}
%access, and improving the design of the (currently very simple) web
%interface with proper usability considerations---along with content
%seeding.

%\section*{Acknowledgments}
%
%This research has received funding from the EC FP7 project
%PRECIOUS under grant agreement no. 611366 and from EIT ICT Labs.
% Seventh Framework Programme
% for research, technological development and demonstration
% (PRECIOUS) and from EIT ICT Labs.

% \footnotesize
%\scriptsize
\small
\bibliographystyle{abbrv}
%\bibliography{../bib/dtn}
\bibliography{dtn}

\end{document}